\documentclass[10pt,aps,prd,twocolumn,notitlepage,showpacs,superscriptaddress]{revtex4-1}
\usepackage{amsthm,amsmath,amsfonts,amssymb,verbatim,color}
\usepackage{graphicx}
\usepackage{bm}
\usepackage{epsfig,slashed,feynmp-auto}
\begin{document}
\begin{fmffile}{diags}

\newcommand{\tr}{\mathop{\mathrm{Tr}}}
\newcommand{\bsigma}{\boldsymbol{\sigma}}
\newcommand{\re}{\mathop{\mathrm{Re}}}
\newcommand{\im}{\mathop{\mathrm{Im}}}
\renewcommand{\b}[1]{{\boldsymbol{#1}}}
\newcommand{\diag}{\mathrm{diag}}
\newcommand{\sign}{\mathrm{sign}}
\newcommand{\sgn}{\mathop{\mathrm{sgn}}}
\renewcommand{\c}[1]{\mathcal{#1}}
\renewcommand{\mathbf}[1]{{\boldsymbol{#1}}}

\newcommand{\mb}{\bm}
\newcommand{\ua}{\uparrow}
\newcommand{\da}{\downarrow}
\newcommand{\ra}{\rightarrow}
\newcommand{\la}{\leftarrow}
\newcommand{\mc}{\mathcal}
\newcommand{\bs}{\boldsymbol}
\newcommand{\lra}{\leftrightarrow}
\newcommand{\nn}{\nonumber}
\newcommand{\half}{{\textstyle{\frac{1}{2}}}}
\newcommand{\mf}{\mathfrak}
\newcommand{\MF}{\text{MF}}
\newcommand{\IR}{\text{IR}}
\newcommand{\UV}{\text{UV}}

\newcommand{\<}{\langle}
\renewcommand{\>}{\rangle}
\newtheorem{corollary}{corollary}

\newcommand*{\vcenteredhbox}[1]{\begingroup
\setbox0=\hbox{#1}\parbox{\wd0}{\box0}\endgroup}
\newcommand{\picscalefactor}{0.5}

\DeclareGraphicsExtensions{.png}

\fmfcmd{%
	style_def wiggly_arrow expr p =
		cdraw (wiggly p);
		cfill (arrow p);
	enddef;}

\title{Superconducting quantum criticality of topological surface states at three loops}

\author{Nikolai Zerf}
\affiliation{Institut f\"ur Theoretische Physik, Universit\"at Heidelberg, D-69120 Heidelberg, Germany}
\affiliation{Department of Physics, University of Alberta, Edmonton, Alberta T6G 2E1, Canada}

\author{Chien-Hung Lin}
\affiliation{Department of Physics, University of Alberta, Edmonton, Alberta T6G 2E1, Canada}

\author{Joseph Maciejko}
\affiliation{Department of Physics, University of Alberta, Edmonton, Alberta T6G 2E1, Canada}
\affiliation{Theoretical Physics Institute, University of Alberta, Edmonton, Alberta T6G 2E1, Canada}
\affiliation{Canadian Institute for Advanced Research, Toronto, Ontario M5G 1Z8, Canada}

\date\today

\begin{abstract}
The semimetal-superconductor quantum phase transition on the two-dimensional (2D) surface of a 3D topological insulator is conjectured to exhibit an emergent $\mathcal{N}=2$ supersymmetry, based on a one-loop renormalization group (RG) analysis in the $\epsilon$ expansion. We provide additional support for this conjecture by performing a three-loop RG analysis and showing that the supersymmetric fixed point found at this order survives the extrapolation to 2D. We compute critical exponents to order $\epsilon^3$, obtaining the more accurate value $\nu\approx 0.985$ for the correlation length exponent and confirming that the fermion and boson anomalous dimensions remain unchanged beyond one loop, as expected from non-renormalization theorems in supersymmetric theories. We further couple the system to a dynamical $U(1)$ gauge field, and argue that the transition becomes fluctuation-induced first order in an appropriate type-I regime. We discuss implications of this result for quantum phase transitions between certain symmetry-preserving correlated surface states of 3D topological insulators.
\end{abstract}


\maketitle

\section{Introduction}\label{intro}

Supersymmetry (SUSY), a spacetime symmetry that exchanges fermions and bosons, plays an important role in modern theories of elementary particles such as superstring theory and the Minimal Supersymmetric Standard Model~\cite{DineBook}. In these theories SUSY is either an exact symmetry or it is broken, spontaneously or explicitly. A third possibility is that SUSY could be a symmetry that emerges at long wavelengths and low energies in theories that are not manifestly supersymmetric~\cite{Thomas05,lee2010}, much like Lorentz symmetry emerges at long wavelengths at the critical point of the familiar two-dimensional (2D) Ising model, whose Hamiltonian is defined on a lattice and is thus not manifestly Lorentz invariant~\cite{Sachdev}. Recent developments suggest that SUSY can emerge naturally at certain quantum critical points (QCP) in a variety of quantum many-body systems, including spinless fermions on the honeycomb lattice with $f$-wave pairing instabilities~\cite{Lee07}, interacting ultracold atomic Fermi gases on certain optical lattices~\cite{yu2010}, the surfaces of topological insulators and superconductors/superfluids~\cite{grover2012,Grover14,Lee14}, Dirac and Weyl semimetals with pair-density-wave instabilities~\cite{Jian15}, and chains of Majorana zero modes~\cite{rahmani2015}.

Here we will focus on emergent $\mathcal{N}=2$ SUSY in 2+1 spacetime dimensions, which was conjectured to occur at the semimetal-superconductor QCP of 2D Dirac fermions on the surface of a 3D topological insulator~\cite{grover2012,Grover14,Lee14}. According to this conjecture, the fermionic charge-$e$ Dirac quasiparticles and bosonic charge-$2e$ Cooper pairs become superpartners at the QCP, at asymptotically low energies and long wavelengths; the critical theory is the superconformal fixed point of the Wess-Zumino model in (2+1)D with one chiral multiplet and cubic superpotential~\cite{aharony1997}. The basis for this conjecture is a one-loop perturbative renormalization group (RG) analysis of the Landau-Ginzburg action for interacting Dirac fermions and Cooper pairs near criticality, carried out in $D=4-\epsilon$ spacetime dimensions using the $\epsilon$ expansion~\cite{Thomas05,Lee07} and extrapolated to the physical case of $D=3$, which corresponds to $\epsilon=1$. In this analysis, the SUSY QCP corresponds to a fixed point of the one-loop RG beta function with a single relevant direction in the infrared corresponding to the coupling constant that tunes the transition, e.g., the strength of the attractive interaction between Dirac fermions.

In this paper we expand upon these previous studies in two directions: (a) we go beyond leading order in the $\epsilon$ expansion, and (b) we investigate the effect of coupling the Dirac fermions and Cooper pairs to a dynamical $U(1)$ gauge field. The motivation for (a) is twofold. First, while there is no reason to doubt the existence of the SUSY fixed point found at one-loop order for infinitesimal $\epsilon$~\cite{Lee07}, the physical case of two spatial dimensions corresponds to $\epsilon=1$. As $\epsilon$ increases from zero to one, higher-order terms in the RG beta function, which correspond to higher loops, become increasingly important and may qualitatively change the fixed-point structure of the theory. For example, the full beta function, if computed to all orders, could have a lower critical spacetime dimension $D_c$ below which the SUSY fixed point disappears. This would correspond to a finite upper critical value $\epsilon_c$, and the disappearance of the fixed point would not be apparent at the lowest orders in perturbation theory. In fact, even if one starts with the manifestly supersymmetric action for the $\mathcal{N}=2$ Wess-Zumino model in (2+1)D, it is strictly speaking not known whether the superconformal fixed point exists in this theory--only that \emph{if} there exists a value of the coupling constant for which the fermion and boson anomalous dimensions are equal to $\frac{1}{3}$, this value corresponds to a fixed point of the RG~\cite{strassler2003}. In general, when the beta function is a function of some external parameter, such as here $\epsilon$, an infrared stable fixed point can disappear at a critical value of this parameter by either merging with the trivial (Gaussian) fixed point, running off to infinite coupling, or merging with an ultraviolet stable fixed point~\cite{kaplan2009}. An example of the latter phenomenon with current relevance to condensed matter physics is the disappearance of the Luttinger-Abrikosov-Beneslavskii non-Fermi liquid fixed point~\cite{moon2013} below a critical spatial dimension $d_c\approx 3.26$ due to its merging with a QCP towards the formation of a topological Mott insulator~\cite{herbut2014}. While only knowing the full non-perturbative beta function of the Landau-Ginzburg theory considered here would establish the existence of the SUSY fixed point in 2+1 dimensions beyond any doubt, such an analysis is likely to be practically impossible, and we limit ourselves to exploring the fixed-point structure of the theory up to three-loop order.

If the SUSY fixed point is found to remain stable at this order all the way down to 2+1 dimensions, our second motivation is to provide more accurate values of universal critical properties. While certain such properties can be determined exactly from SUSY, such as the fermion and boson anomalous dimensions~\cite{aharony1997} and the universal zero-temperature critical conductivity~\cite{witczak-krempa2015}, the correlation length exponent $\nu$---and other critical exponents that depend on it via scaling laws---are not fixed exactly by SUSY and have been determined only to leading order in the $\epsilon$ expansion. Our three-loop RG analysis allows us to calculate $\nu$ to order $\epsilon^3$ [Eq.~(\ref{nu3loop})], giving $\nu\approx 0.985$ in 2+1 dimensions. This is a 7\% difference with the conformal bootstrap result $\nu\approx 0.917$~\cite{Bobev}, compared with 18\% for the order-$\epsilon$ result $\nu\approx 0.75$~\cite{Thomas05,Lee07}. While the computation of $\nu$ typically requires renormalizing a massive theory, as in the case of the $O(N)$ Wilson-Fisher fixed point~\cite{Sachdev}, here we give the explicit proof of an exact relation~\cite{Thomas05} arising from SUSY between $\nu$ and the stability critical exponent $\omega$ of the massless Wess-Zumino model that allows us to calculate $\nu$ in the massless theory, i.e., the Landau-Ginzburg theory restricted to its critical hypersurface.

Finally, the motivation for studying the effect on the QCP of coupling the system to a dynamical $U(1)$ gauge field is also twofold. First, an interesting result was reported in Ref.~\cite{roy2013}, namely that in the $\epsilon$ expansion below four dimensions the transition remains continuous in the presence of such a coupling. This would be in stark contrast with the purely bosonic case where the superconducting transition becomes fluctuation-induced first order~\cite{halperin1974,coleman1973} in the $\epsilon$ expansion. Second, the problem of a single species of Dirac fermions coupled to a dynamical $U(1)$ gauge field in (2+1)D, i.e., three-dimensional quantum electrodynamics (QED$_3$), has received much attention lately as a possible dual description of the topological surface states~\cite{Wangduality,Metlitskiduality,wang2016,mross2015,karch2016,seiberg2016,murugan2016}. The superconducting transition of Dirac fermions in the dual QED$_3$ corresponds in the original theory to a transition from the dual Dirac liquid to the T-Pfaffian state~\cite{Wangduality}, two possible strongly correlated surface states of the 3D topological insulator. Under this conjectured duality the time-reversal ($T$) symmetry of the original topological insulator is mapped to particle-hole/charge-conjugation symmetry in the dual theory. The dual Dirac fermions undergoing a pairing transition must therefore be at zero chemical potential, as we consider here, rather than in the generic situation of being paired on a finite helical Fermi surface~\cite{fu2008}. Studying the effect of a dynamical $U(1)$ gauge field on the SUSY QCP of superconducting Dirac fermions should thus shed light on the nature of the transition between the dual Dirac liquid and the T-Pfaffian state. We revisit this problem and find, in disagreement with Ref.~\cite{roy2013}, that the QCP disappears in the presence of the gauge field; we interpret this as meaning that the superconducting transition becomes first order, as in the bosonic problem. At the level of the $\epsilon$ expansion, the transition between the dual Dirac liquid and the T-Pfaffian state is thus fluctuation-induced first order; we conjecture this result to hold in (2+1)D in an appropriate type-I regime for the transition, as in the bosonic case~\cite{kleinert1982}. Of course, the bosonic transition is known to become continuous in the type-II regime~\cite{dasgupta1981,kleinert1982}, and an analogous regime could exist for the correlated topological surface states as well.

The rest of the paper is organized as follows. In Sec.~\ref{section_rg}, we review one-loop RG results for the semimetal-superconductor QCP on the surface of a 3D topological insulator. We then show that at three-loop order the critical spacetime dimension below which the SUSY QCP disappears is $D_c\approx 1.62$.  Thus, at three-loop order the SUSY fixed point remains stable all the way down to 2+1 spacetime dimensions. In Sec.~\ref{section_exp}, we compute various critical exponents. In Sec.~\ref{section_gaugefield}, we consider the critical theory coupled to a $U(1)$ gauge field. We argue by a combination of mean-field and RG arguments that in this case, the phase transition becomes fluctuation-induced first order. Finally, we conclude in Sec.~\ref{section_conclusion}. Technical details can be found in Appendix~\ref{app_rgconstant}-\ref{app_rg}.

\section{Renormalization group analysis \label{section_rg}}

While $\mathcal{N}=2$ SUSY has been proposed to emerge in a variety of physical systems, for definiteness we focus here on its potential realization on the 2D surface of a 3D $T$-invariant topological insulator~\cite{hasan2010,qi2011} with a single gapless two-component Dirac fermion $\psi=(\psi_\ua,\psi_\da)$ on the surface. For weak electron-electron interactions, the gaplessness of the Dirac surface state is protected by $T$ symmetry and $U(1)$ particle-number conservation symmetry. We assume that the chemical potential is at the Dirac point, and consider the scenario in which interactions are sufficiently strong and attractive to spontaneously break the $U(1)$ symmetry and induce superconductivity on the surface~\cite{roy2013,nandkishore2013}. The superconducting phase is characterized by a nonzero expectation value of the Cooper pair bosonic order parameter $\phi$. The imaginary-time Landau-Ginzburg Lagrangian governing the semimetal-superconductor quantum phase transition is
\begin{equation}
\mathcal{L}=i\bar{\psi}\slashed{\partial}\psi+|\partial_{\mu}\phi|^2+m^2|\phi|^{2}+\lambda^2|\phi|^{4}
+h(\phi^*\psi^{T}i\sigma_{2}\psi+\mathrm{h.c.}),\label{lti}%
\end{equation}
where $\bar{\psi}=-i\psi^\dag\gamma_0$ and $\slashed{\partial}=\gamma_\mu\partial_\mu$, and we define $\gamma_0=\sigma_3$, $\gamma_1=\sigma_1$, and $\gamma_2=\sigma_2$ the Pauli matrices. Note that a fermion mass term $\propto\bar{\psi}\psi$ is forbidden by $T$ symmetry. The semimetal with $\langle\phi\rangle=0$ is found for $m^2>0$ and the superconductor with $\langle\phi\rangle\neq 0$ is found for $m^2<0$; the QCP is obtained by tuning $m^2$ to zero where both fermionic and bosonic fields are gapless (here $m$ stands for the renormalized mass). The Lagrangian is invariant under global $U(1)$ transformations $\psi\rightarrow e^{i\theta}\psi$, $\phi\rightarrow e^{2i\theta}\phi$, implying that $\phi$ carries twice the charge of $\psi$. The Lagrangian is also Lorentz invariant; while this is not an exact symmetry in general, a one-loop RG analysis of a Lorentz-breaking version of theory (\ref{lti}) with different fermion and boson velocities shows that these two velocities flow to the same value in the low-energy limit~\cite{Lee07,roy2016}. In other words, Lorentz invariance emerges in that limit. This is independent of the couplings $h$ (assuming it is nonzero), $m^2$, and $\lambda^2$, and of the value of $\epsilon$. We thus expect it to be a robust feature of the semimetal-superconductor transition for Dirac fermions regardless of the nature of the QCP~\cite{roy2016}, and we assume Lorentz invariance from the outset.

\subsection{Review of the one-loop results\label{section_1loop}}

The critical properties of the Landau-Ginzburg theory (\ref{lti}) have been investigated previously using one-loop RG in $D=4-\epsilon$ spacetime dimensions on the critical hypersurface $m=0$. The physical case of (2+1)D Dirac fermions corresponds to $\epsilon=1$. The (ultraviolet) one-loop beta functions $\beta_{h^2}=dh^2/d\ln\mu$ and $\beta_{\lambda^2}=d\lambda^2/d\ln\mu$ for the rescaled couplings $\frac{h^{2}}{(4\pi)^2}\rightarrow h^2$ and $\frac{\lambda^{2}}{(4\pi)^2}\rightarrow\lambda^2$ are~\cite{Thomas05,Lee07}
\begin{align}
	\beta_{h^2} &= -\epsilon h^2+12 h^4,\label{1Lbeta1}\\
	\beta_{\lambda^2} &=-\epsilon \lambda^2+20 \lambda^4 +8h^2\lambda^2 -16h^4.\label{1Lbeta2}
\end{align}
Here $\mu$ is the energy scale parameterizing the flow of couplings.
To leading order in the $\epsilon$ expansion, there are two unstable fixed points, the Gaussian fixed point $(h_*^2,\lambda_*^2)=(0,0)$ and the $O(2)$ Wilson-Fisher fixed point $(h_*^2,\lambda_*^2)=(0,\frac{\epsilon}{20})$, and one stable fixed point $(h_*^2,\lambda_*^2)=(\frac{\epsilon}{12},\frac{\epsilon}{12})$. To this order in perturbation theory, the latter fixed point remains stable as one extrapolates $\epsilon\rightarrow 1$. As this fixed point has only one relevant direction corresponding to the tuning parameter $m^2$, it is identified with the semimetal-superconductor QCP. $h_*^2=\lambda_*^2$ implies that the critical theory is invariant under the $\mathcal{N}=2$ SUSY transformations~\cite{Lee07}
\begin{align}
\delta_\eta\phi&=-\bar{\psi}\eta, &&\delta_\eta\phi^*=\bar{\eta}\psi, \nn\\
\delta_\eta\psi&=i\slashed{\partial}\phi^*\eta-\frac{h}{2}\phi^2i\sigma_2\bar{\eta}^T,
&&\delta_\eta\bar{\psi}=i\bar{\eta}\slashed{\partial}\phi-\frac{h}{2}\phi^{*2}\eta^Ti\sigma_2,
\end{align}
where $\eta$ is a two-component Grassmann spinor that parametrizes the transformation. The QCP thus has emergent $\mathcal{N}=2$ SUSY at the one-loop level.

\subsection{Three-loop analysis\label{section_3loop}}

As argued in Sec.~\ref{intro}, an important question is the stability of the SUSY fixed point against higher order terms in the RG beta function. We attack this problem by performing a three-loop RG analysis in $D=4-\epsilon$ spacetime dimensions on the quantum critical hypersurface $m=0$. We first introduce some notations. For the RG procedure, we interpret Eq.~(\ref{lti}) as a bare Lagrangian with bare fields and bare couplings that we denote $\psi_0$, $\phi_0$ and $\lambda_0$, $h_0$, respectively. We denote the corresponding renormalized fields and couplings by $\psi$, $\phi$, $\lambda$, and $h$. The renormalized Lagrangian is
\begin{eqnarray}
	\mathcal{L}&=&iZ_\psi\bar{\psi}\slashed{\partial}\psi+Z_\phi|\partial_{\mu}\phi|^2+Z_\lambda^2 \lambda^2	\mu^\epsilon|\phi|^{4} \nonumber\\
	&&+Z_h h\mu^{\epsilon/2}(\phi^*\psi^{T}i\sigma_{2}\psi+\mathrm{h.c.}),
	\label{lti2}
\end{eqnarray}
where we define wave function renormalization constants $Z_\psi$ and $Z_\phi$ and vertex renormalization constants $Z_\lambda$ and $Z_h$. The renormalized Lagrangian can be obtained from the bare one by rescaling the fields,
\begin{align}
	\psi_0=\sqrt{Z_\psi}\psi,\ \ \ 
	\phi_0=\sqrt{Z_\phi}\phi,\label{wfrgconstant}
\end{align}
which implies the two relations
\begin{align}
	h^2&=h_0^2 \mu^{-\epsilon} Z_\psi^2 Z_\phi Z_h^{-2}	, \label{couplingsh}\\
	\lambda^2&=\lambda_0^2 \mu^{-\epsilon} Z_\phi^2 Z_\lambda^{-2}.
	\label{couplingslambda}
\end{align}
Because the bare couplings are independent of $\mu$, the relations (\ref{couplingsh})-(\ref{couplingslambda}) describe the scale dependence of the renormalized couplings. We compute the wave function renormalization constants $Z_\psi$ and $Z_\phi$ from the self-energy diagrams of the fermionic and bosonic fields at three-loop order using dimensional regularization and the modified minimal subtraction scheme ($\overline{\text{MS}}$). The vertex renormalization constants $Z_\lambda$ and $Z_h$ are computed from the $|\phi|^4$ and fermion-boson vertex diagrams to the same order. Specifically, we expand all renormalization constants $Z_x$ with $x \in \{\psi,\phi,\lambda,h\}$ in the form
\begin{equation}
	Z_x(h^2,\lambda^2)=1+\sum_{L=1}^{N_\text{loops}} \delta Z_{x,L}(h^2,\lambda^2),
	\label{}
\end{equation}
where 1 is the tree-level term and the loop correction $\delta Z_{x,L}(h^2,\lambda^2)$ is decomposed into products of coupling constants
\begin{equation}
	\delta Z_{x,L}(h^2,\lambda^2)=\sum_{i+j=L}(h^2)^i(\lambda^2)^j\delta Z_{x,L}(i,j).
	\label{}
\end{equation}
Here $L$ denotes the order of the expansion and the coefficients $\delta Z_{x,L}(i,j)$ contain poles of maximum order $L$ at $\epsilon=0$. The renormalization constants at three-loop order ($N_\text{loops}=3$) are listed in Appendix~{\ref{app_rgconstant}}.

\begin{figure}
[t]
\begin{center}
\includegraphics[width=0.7\columnwidth]%
{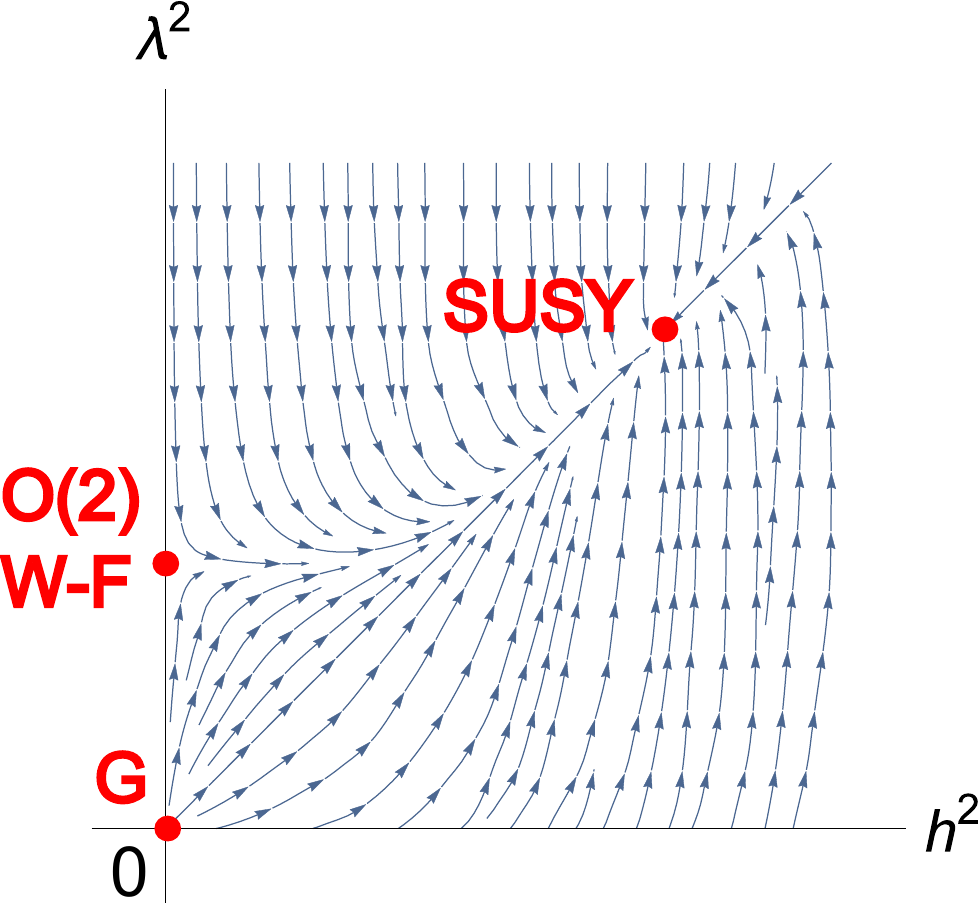}%
\caption{Three-loop RG flow of the Landau-Ginzburg Lagrangian (\ref{lti}) on the quantum critical hypersurface, with the Gaussian (G), $O(2)$ Wilson-Fisher (W-F), and SUSY fixed points.}
\label{fig:rgflow}
\end{center}
\end{figure}

Having obtained these renormalization constants, the beta functions for the coupling constants $h^2$ and $\lambda^2$ can be computed from Eq.~(\ref{couplingsh})-(\ref{couplingslambda}) by taking a derivative with respect to $\ln\mu$ on both sides of the equations, remembering that the bare couplings $h_0^2$, $\lambda_0^2$ are independent of $\mu$:
\begin{align}
	\beta_{h^2}&=\left(-\epsilon+2\gamma_\psi+\gamma_\phi-2\gamma_h\right)h^2, \label{betah2}\\
	\beta_{\lambda^2}&=\left(-\epsilon+2\gamma_\phi-2\gamma_\lambda\right)\lambda^2, \label{betal2}
\end{align}
where we define
\begin{align}\label{gamma}
\gamma_x=\frac{d\ln Z_x}{d\ln\mu}=\beta_{h^2}\frac{\partial\ln Z_x}{\partial h^2}+\beta_{\lambda^2}\frac{\partial\ln Z_x}{\partial\lambda^2},
\end{align}
for $x\in\{\psi,\phi,\lambda,h\}$, since the renormalization constants are functions of the (renormalized) coupling constants. Equations (\ref{betah2})-(\ref{gamma}) thus form a linear system of equations for the beta functions that can be solved as
\begin{align}
\left(\begin{array}{c}
		\beta_{h^2}\\
		\beta_{\lambda^2}
	\end{array}\right)
	=\epsilon M^{-1}
	\left(\begin{array}{c}
		h^2 \\
		\lambda^2
	\end{array}\right),
\end{align}
where
\begin{align}
M_{11}&=-1-h^2\left(2\frac{\partial\ln Z_\psi}{\partial h^2}+\frac{\partial\ln Z_\phi}{\partial h^2}-2\frac{\partial\ln Z_h}{\partial h^2}\right),\\
M_{12}&=-h^2\left(2\frac{\partial\ln Z_\psi}{\partial\lambda^2}+\frac{\partial\ln Z_\phi}{\partial\lambda^2}-2\frac{\partial\ln Z_h}{\partial\lambda^2}\right),\\
M_{21}&=-\lambda^2\left(2\frac{\partial\ln Z_\phi}{\partial h^2}-2\frac{\partial\ln Z_\lambda}{\partial h^2}\right),\\
M_{22}&=-1-\lambda^2\left(2\frac{\partial\ln Z_\phi}{\partial\lambda^2}-2\frac{\partial\ln Z_\lambda}{\partial\lambda^2}\right).
\end{align}
So far the beta functions are exact in terms of renormalization constants. In practice, we can only compute these renormalization constants to finite order in perturbation theory and we obtain the beta functions at the same order as the renormalization constants. Specifically, by inserting the renormalization constants $Z_\psi$, $Z_\phi$, $Z_h$, and $Z_\lambda$ in the matrix $M$ and expanding in powers of the coupling constants, we obtain the beta functions to the desired order in the coupling constants $h^2$ and $\lambda^2$. Using the three-loop renormalization constants (see Appendix~\ref{app_rgconstant}), we obtain the three-loop beta functions:
\begin{eqnarray}
	\beta_{h^2}  & =&-\epsilon h^{2}+12h^{4}+8h^{2}\lambda^{4}-64h^{4}\lambda^{2}+8h^{6}-40h^{2}\lambda^{6}\nn\\
	&&+300h^{4}\lambda^{4}+1632h^{6}\lambda^{2}-[1652- 576\zeta(3)]  h^{8},
\label{beta1}\\
\beta_{\lambda^2}  & =&-\epsilon\lambda^{2}+20\lambda^{4}+8h^{2}\lambda^{2}-16h^{4}-80h^{2}\lambda^{4}+16h^{4}\lambda^{2}\nn\\
&&+256h^{6}-240\lambda^{6}+904h^{2}\lambda^{6}\nn\\
&&+[3832+3264\zeta(3)] h^{4}\lambda^{4}-[8664+2688\zeta(3)]h^{6}\lambda^{2}\nn\\
&&-[768+ 3072\zeta(3)] h^{8}+[4936+3072\zeta(3)]\lambda^{8},
\label{beta2}
\end{eqnarray}
where $\zeta(3)=1.2020569\ldots$ is Ap\'{e}ry's constant, with $\zeta(z)$ the Riemann zeta function. Two limiting cases previously considered in the literature can be recovered. Setting $h=0$, the free Dirac fermion $\psi$ and the complex scalar $\phi$ decouple, and Eq.~(\ref{beta2}) reproduces the three-loop beta function for the bosonic $O(2)$ vector model~\cite{Justinbook}. If one sets $h=\lambda$, the bare theory has exact (rather than emergent) $\mathcal{N}=2$ SUSY; Eq.~(\ref{beta1}) and (\ref{beta2}) become equal and reproduce the three-loop beta function for the Wess-Zumino model~\cite{avdeev1982}.

\begin{figure}
[t]
\begin{center}
\includegraphics[width=0.9\columnwidth]%
{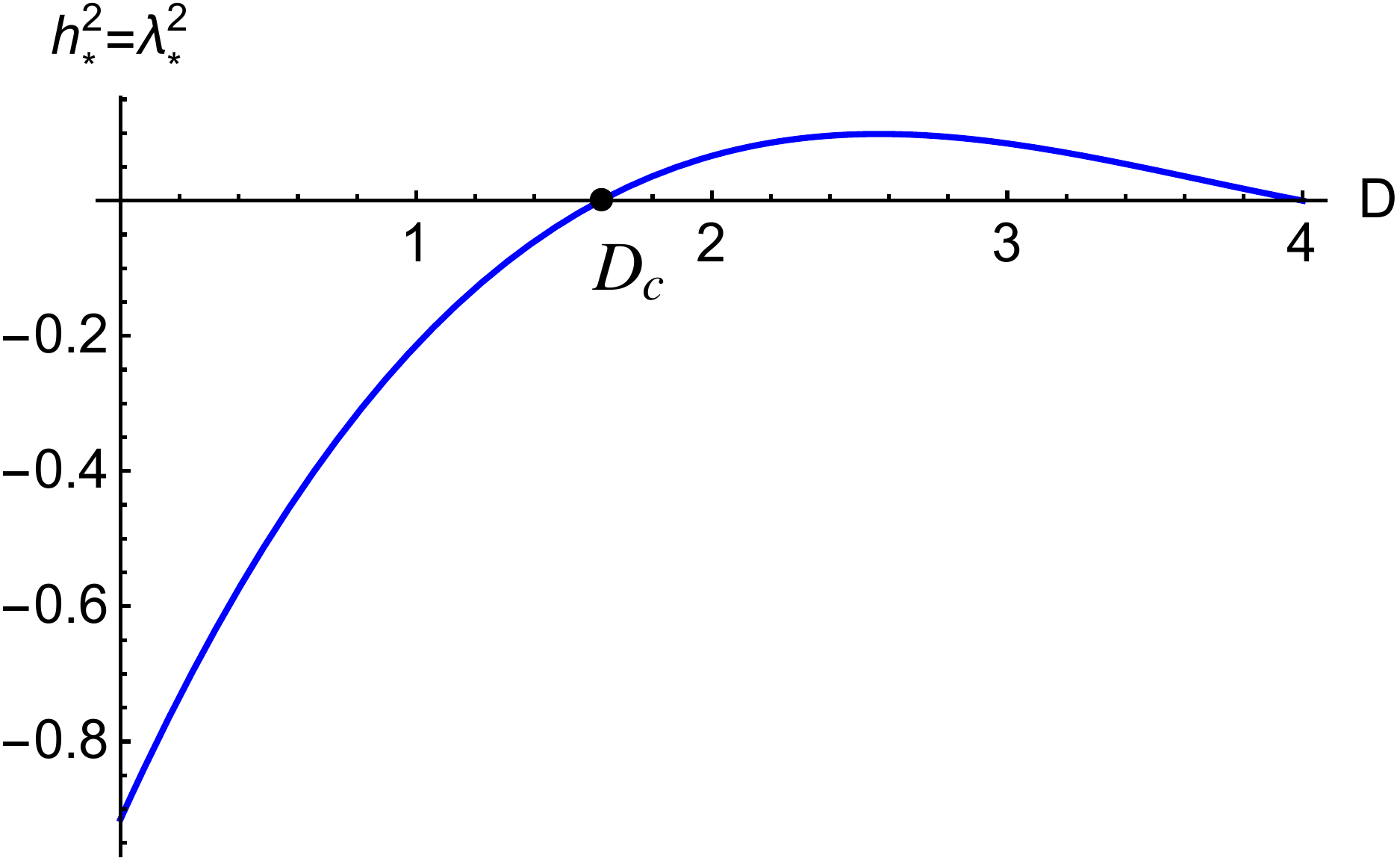}%
\caption{Couplings at the SUSY fixed point as a function of spacetime dimension $D$ from the three-loop RG beta function. At this order, the critical dimension $D_c\approx 1.62$ is still well below the physical dimension $D=3$.}
\label{fig:critdim}
\end{center}
\end{figure}

The fixed points of the RG flow can be solved for order by order in $\epsilon$ (Fig.~\ref{fig:rgflow}). We find two infrared unstable fixed points: the Gaussian and $O(2)$ Wilson-Fisher fixed points,
\begin{align}
	\left(  h^{2}_*,\lambda^{2}_*\right)  &=\left(  0,0\right),\\
	\left(  h^{2}_*,\lambda^{2}_*\right)  &= \left(  0,\frac{\epsilon}{20}  +\frac{3 \epsilon^{2}}{100} -
	\frac{384 \zeta(3)-103}{20000} \epsilon^3 \right),
\end{align}
and the infrared stable fixed point with
\begin{equation}
	h^{2}_*=\lambda^{2}_*=\frac{\epsilon}{12}+\frac{\epsilon^{2}}{36}-\frac
{4\zeta(3)-1 }{144}\epsilon^{3}.
	\label{susyfixedpt}
\end{equation} 
At this fixed point, the theory exhibits emergent $\mathcal{N}=2$ SUSY as in the one-loop case (Sec.~\ref{section_1loop}). The critical spacetime dimension below which the fixed point couplings (\ref{susyfixedpt}) become negative is $D_c\approx 1.62$. Thus, at three-loop order the SUSY fixed point in (2+1)D is still in a physical range of coupling constants (Fig.~\ref{fig:critdim}).

\section{Critical exponents \label{section_exp}}

In this section, we compute anomalous dimensions $\gamma_\phi,\gamma_\psi$ for the bosonic and fermionic fields and the correlation length exponent $\nu$. All other critical exponents can then be obtained from $\gamma_\phi,\gamma_\psi,\nu$ by scaling relations.

\subsection{Anomalous dimensions}

The anomalous dimensions $\gamma_\phi^*$ and $\gamma_\psi^*$ for the bosonic field $\phi$ and the fermionic field $\psi$ are obtained by evaluating the logarithmic derivatives of the renormalization constants (\ref{gamma}) at a given fixed point,
\begin{equation}
	\gamma_\phi^*=\gamma_\phi(h^2_*,\lambda^2_*),\ \gamma_\psi^*=\gamma_\psi(h^2_*,\lambda^2_*).
\end{equation}
Using the renormalization constants $Z_\phi$, $Z_\psi$ from Appendix~\ref{app_rgconstant}, we obtain
\begin{eqnarray}
	\gamma_\phi&=&4h^2-24h^4+8\lambda^4-40\lambda^6-60h^2\lambda^4\nn\\
	&&+160h^4\lambda^2+[20+192\zeta(3)]h^6,\\
	\gamma_\psi&=&4h^2-16h^4-44h^2\lambda^4+128h^4\lambda^2+[192\zeta(3)-4]h^6.\nn\\
\end{eqnarray}
To obtain the anomalous dimensions at the SUSY fixed point, we insert Eq.~(\ref{susyfixedpt}) into the above equations and find that the bosonic field $\phi$ and the fermionic field $\psi$ have the same anomalous dimension at three-loop order,
\begin{equation}\label{anomalousdim}
	\gamma_\phi^*=\gamma_\psi^*=\frac{\epsilon}{3} + O(\epsilon^4),
\end{equation}
as expected for a fixed point with emergent SUSY, since all fields within the same supermultiplet should have the same scaling dimensions. Furthermore, the three-loop result (\ref{anomalousdim}) is the same as the one-loop result~\cite{Thomas05,Lee07}, as expected from SUSY non-renormalization theorems~\cite{strassler2003}. The anomalous dimensions have no quantum corrections beyond one loop, that is, the one-loop expansion is exact; the perfect cancellation of terms of order $\epsilon^2$ and $\epsilon^3$ in Eq.~(\ref{anomalousdim}) is thus a strong check on the validity of our three-loop calculation.

\subsection{Correlation length exponent}

To define the correlation length exponent $\nu$, one typically keeps a bare mass term $m_0^2|\phi_0|^2$ in the theory (\ref{lti}). The renormalized theory (\ref{lti2}) then contains a renormalized mass term $Z_{m^2}m^2\mu^2|\phi|^2$, with a renormalization constant $Z_{m^2}$ and dimensionless renormalized mass $m^2$. Setting the bare and renormalized mass terms equal to each other, we obtain the relation $m^2=m_0^2\mu^{-2}Z_\phi Z_{m^2}^{-1}$ between bare and renormalized masses, from which the RG beta function for the mass can be obtained as
\begin{align}\label{betam2_maintext}
	\beta_{m^2}=\left(-2+\gamma_\phi-\gamma_{m^2}\right)m^2, 
\end{align}
where $\gamma_\phi=d\ln Z_\phi/d\ln\mu$ and $\gamma_{m^2}=d\ln Z_{m^2}/d\ln\mu$ [see Eq.~(\ref{gamma})]. The correlation length exponent $\nu$ is then defined as minus the inverse of the coefficient of $m^2$ in the beta function (\ref{betam2_maintext}), evaluated at the fixed point of interest~\cite{Justinbook},
\begin{align}\label{nu}
\nu^{-1}=2-\gamma_\phi^*+\gamma_{m^2}^*.
\end{align}
Using this method~\cite{roy2013,nandkishore2013}, $\nu$ for the SUSY QCP considered here can be evaluated at one-loop order,
\begin{align}\label{nuoneloop}
\nu=\frac{1}{2}+\frac{\epsilon}{4}+\mathcal{O}(\epsilon^2),
\end{align}
which gives $\nu\approx 0.75$ in (2+1)D, corresponding to a 18\% difference with the conformal bootstrap result $\nu\approx 0.917$~\cite{Bobev}. To calculate $\nu$ at three-loop order and thus obtain a better estimate of its value in (2+1)D, we would in principle have to renormalize the massive theory (\ref{lti}) at three-loop order, which is technically difficult. However, the one-loop result (\ref{nuoneloop}) was in fact obtained originally~\cite{Thomas05,Lee07} not by renormalizing the massive theory but by exploiting a nontrivial consequence of $\mathcal{N}=2$ SUSY, namely, the exact relation~\cite{Thomas05}
\begin{align}\label{theo}
	\omega=\gamma_{\phi}^*-\gamma_{m^2}^*,
\end{align}
where
\begin{align}\label{omegadef}
\omega\equiv\frac{d\beta_{h^2}(h_*^2)}{dh^2},
\end{align}
is the slope of the beta function in the massless theory (i.e., on the critical hypersurface), evaluated at the SUSY fixed point. We define $\beta_{h^2}(h^2)\equiv\left.\beta_{h^2}(h^2,\lambda^2)\right|_{\lambda^2=h^2}$ as the beta function along a SUSY-preserving trajectory. The exponent $\omega$, sometimes known as the stability critical exponent, characterizes the approach to the fixed point ($\omega>0$ means the fixed point is infrared stable with respect to the coupling $h^2$, while $\omega<0$ means it is unstable) as well as leading corrections to scaling behavior in the critical region~\cite{Justinbook}. Putting Eq.~(\ref{theo}) and (\ref{nu}) together, we obtain~\cite{Thomas05}
\begin{align}\label{exactrelation}
\nu^{-1}=2-\omega.
\end{align}
Knowing the three-loop beta functions (\ref{beta1})-(\ref{beta2}) in the massless theory, we can thus use Eq.~(\ref{omegadef}) and (\ref{exactrelation}) to calculate $\nu$ at three-loop order without having to renormalize the massive theory.

Let us evaluate the correlation length exponent for our theory (\ref{lti}) at the SUSY fixed point by using the exact relation (\ref{exactrelation}).
The stability critical exponent at the SUSY fixed point can be computed from Eq.~(\ref{beta1}), (\ref{susyfixedpt}), and (\ref{omegadef}) at three-loop order:
\begin{equation}
	\omega=\epsilon-\frac{\epsilon^2}{3}+\left(\frac{1}{18}+\frac{2\zeta(3)}{3}\right)\epsilon^3+O(\epsilon^4).
	\label{omega}
\end{equation}
Thus, the correlation length exponent $\nu$ can be obtained from Eq.~(\ref{exactrelation}) at the same order:
\begin{equation}
	\nu=\frac{1}{2}+\frac{\epsilon}{4}+\frac{\epsilon^2}{24}+\left(\frac{\zeta(3)}{6}-\frac{1}{144}\right)\epsilon^3+O(\epsilon^4).
	\label{nu3loop}
\end{equation}
In 2+1 dimensions (i.e., $\epsilon=1$), Eq.~(\ref{nu3loop}) gives $\nu\approx 0.985$. This is a 7\% difference with the conformal bootstrap result $\nu\approx 0.917$~\cite{Bobev}, compared with 18\% for the one-loop result (\ref{nuoneloop}).

The derivation of Eq.~(\ref{theo}) was briefly sketched in Ref.~\cite{Thomas05}; for completeness we here provide an explicit, detailed derivation. The basic idea is to use the superspace formalism of SUSY~\cite{DineBook} to show that the mass term $m^2|\phi|^2$ in the Lagrangian (\ref{lti}) can be absorbed by a Grassmann-valued rescaling of the chiral superfield $\Phi$ (which contains both bosonic $\phi$ and fermionic $\psi$ components), allowing one to relate properties of the massive, SUSY-breaking theory to those of the massless, SUSY-preserving Wess-Zumino model. In the context of particle physics, the scalar mass term $m^2|\phi|^2$ is referred to as a soft breaking of SUSY, and the approach described above allows one to calculate the RG beta function $\beta_{m^2}$ of the mass in a softly broken SUSY theory~\cite{Avdeev}. This is precisely what we need, given that the correlation length exponent $\nu$ is related to $\beta_{m^2}$ [see Eq.~(\ref{betam2_maintext})-(\ref{nu})].

Our starting point is the (bare) Lagrangian (\ref{lti}) on a SUSY-preserving trajectory $\lambda^2=h^2$, $m^2=0$ and written in the superspace notation (see Appendix~\ref{app_superspace}),
\begin{equation}
	\mathcal{L}_\text{rigid}=\int d^{2}\theta d^{2}\bar{\theta}\,\Phi_0^{\dagger}\Phi_0+\left(  \int
d^{2}\theta\frac{h_0}{3}\Phi_0^{3}+\text{h.c.}\right),  \label{rigid}%
\end{equation}
where $\Phi_0$ is the chiral superfield. We use the subscript $0$ to denote bare quantities. Following the particle physics terminology we will call Eq.~(\ref{rigid}) the rigid SUSY theory, by contrast with the softly broken SUSY theory with $\lambda_0^2=h_0^2$ but $m_0^2\neq 0$. When we expand $\Phi_0$ into its components $\psi_0$, $\phi_0$, and the auxiliary field $F_0$, integrate over the Grassmann coordinates $\theta,\bar{\theta}$, and integrate out $F_0$, the first term of Eq.~(\ref{rigid}) yields the kinetic terms for $\psi_0$ and $\phi_0$, and the superpotential $\propto\Phi_0^3$ gives the fermion-boson interaction and boson self-interaction terms. In order to derive Eq.~(\ref{theo}), we add to the rigid theory the boson mass term,
\begin{align}
	\mathcal{L}_\text{soft-breaking}=m_0^{2}|\phi_0|^2.
\end{align}
Remarkably, in the superspace formalism the full Lagrangian $\mathcal{L}_\text{soft}=\mathcal{L}_\text{rigid}+\mathcal{L}_\text{soft-breaking}$ for the softly broken theory can be written as%
\begin{eqnarray}
	\mathcal{L}_\text{soft}&=&\int d^{2}\theta d^{2}\bar{\theta}\,\Phi_0^{\dagger}(1+m_0^{2}\theta^{2}\bar{\theta}^{2})\Phi_0\nonumber\\
	&&+\left(  \int d^{2}\theta\frac{h_0}{3} \Phi_0^{3}+\text{h.c.} \right),
	\label{soft}
\end{eqnarray}
i.e., the soft SUSY-breaking mass term enters as a Grassmann-valued multiplicative correction to the kinetic term $\Phi_0^\dag\Phi_0$ of the rigid theory.

We can rewrite $\mathcal{L}_{\text{rigid}}$ in terms of a renormalized superfield $\Phi$, wave function renormalization constant $Z$, and dimensionless renormalized coupling $h$,
\begin{align}
\mathcal{L}_\text{rigid}=\int d^{2}\theta d^{2}\bar{\theta}Z\Phi^{\dagger}\Phi +\left(  \int d^{2}\theta\frac{h\mu^{\epsilon/2}}{3}  \Phi^{3}+\text{h.c.}\right). \label{rigid1}
\end{align}
Non-renormalization theorems ensure there is no renormalization of the superpotential~\cite{grisaru1979,seiberg1993}, such that the corresponding renormalization constant $Z_h$ is equal to one at all orders in perturbation theory: we have checked this explicitly at three-loop order (see Appendix~\ref{app_rgconstant}). By contrast with the rigid theory, for the softly broken theory $\mathcal{L}_{\text{soft}}$ the wave function renormalization constant is promoted to a non-dynamical superfield $\tilde{Z}$,
\begin{eqnarray}
\mathcal{L}_\text{soft}&=&\int d^{2}\theta d^{2}\bar{\theta} \tilde{Z}\Phi^{\dagger}(1+m^2\mu^2 \theta^{2}\bar{\theta}^{2})\Phi  \nonumber\\
	&&+\left(  \int d^{2}\theta\frac{h\mu^{\epsilon/2}}{3}  \Phi^{3}+\text{h.c.}\right),\label{soft1}
\end{eqnarray}
where we have introduced a dimensionless renormalized mass $m$.  To proceed with the renormalization of the softly broken theory, we use the fact that the renormalization superfield $\tilde{Z}$ of the softly broken theory and the renormalization constant $Z$ of the rigid theory are related by the equality~\cite{Avdeev}
\begin{equation}
	\tilde{Z}(h)=Z(\tilde{h}),
	\label{zrelation}
\end{equation}
with 
\begin{equation}
	\tilde{h}=h-\frac{3}{2}m^2 \mu^2 h\theta^2\bar{\theta}^2.
	\label{}
\end{equation}
Then, by expanding the right-hand side of Eq. (\ref{zrelation}) in powers of the Grassmann coordinates $\theta,\bar{\theta}$, we obtain the desired relation between $\tilde{Z}$ and $Z$:
\begin{equation}
	\tilde{Z}(h)=Z(h)\left(1-\frac{3}{2} m^2 \mu^2 h\frac{d\ln Z(h)}{dh} \theta^2\bar{\theta}^2\right).
	\label{zrelation1}
\end{equation}
One way to derive Eq.~(\ref{zrelation}) is to rescale the superfield $\Phi \ra \left(1-\frac{1}{2}m^2\mu^2\theta^2\bar{\theta}^2\right)\Phi$ in Eq.~(\ref{soft1}) and rewrite the latter as
\begin{equation}
	\mathcal{L}_\text{soft}=\int d^{2}\theta d^{2}\bar{\theta}\tilde{Z}\Phi^{\dagger}\Phi +\left(  \int d^{2}\theta\frac{\tilde{h}\mu^{\epsilon/2}}{3}  \Phi^{3}+\text{h.c.}\right) .
	\label{soft2}
\end{equation}
Now $\mathcal{L}_{\text{soft}}$ in Eq.~(\ref{soft2}) looks exactly like $\mathcal{L}_{\text{rigid}}$ in Eq.~(\ref{rigid1}) with the replacement $Z\ra\tilde{Z}$ and $h\ra\tilde{h}$. Accordingly, we deduce the equality (\ref{zrelation}). 

With Eq.~(\ref{zrelation}) at hand, we follow the same procedure as in Sec.~\ref{section_rg} to compute the RG beta functions. The wave function renormalization constant $Z(h)$ relates the bare and renormalized superfields,
\begin{equation}
	\Phi_0=\sqrt{Z(h)}\Phi.
	\label{}
\end{equation}
Inserting this relation in Eq.~(\ref{soft}) and comparing with Eq.~(\ref{soft1}), we can read out the two relations
\begin{align}
	m^2&=m_0^2\mu^{-2}\left(1-\frac{3}{2}h\frac{d\ln Z(h)}{dh}\right)^{-1}, \label{m2eq}\\
	h^2&=h_0^2\mu^{-\epsilon}Z(h)^{3}.
\end{align}
Next, we take a derivative with respect to $\ln\mu$ on both sides of the equations and obtain the beta functions,
\begin{align}
	\beta_{m^2}&=\left(-2+3h^2\frac{d\gamma(h)}{dh^2}\right)m^2,\label{betam2sf}\\
	\beta_{h^2}&=\left(-\epsilon+3\gamma(h)\right)h^2, \label{betah2sf}
\end{align}
where we define
\begin{equation}
	\gamma(h)=\frac{d\ln Z(h)}{d \ln\mu}.
	\label{}
\end{equation}

We can now derive the relation (\ref{theo}) at the SUSY fixed point $h=h_*$ where we have $m=0$ and $\gamma(h_*)=\epsilon/3$.
On the one hand, by comparing Eq.~(\ref{betam2sf}) and Eq.~(\ref{betam2_maintext}), we can identify
\begin{align}
	\gamma_{\phi}^*-\gamma_{m^2}^*=3h_*^2\frac{d\gamma(h_*)}{dh^2}.
\end{align}
On the other hand, $\omega$ can be computed from Eq.~(\ref{betah2sf}),
\begin{align}
\omega\equiv\frac{d\beta_{h^{2}}(h_*^2)}{d h^{2}}=
3h_*^2\frac{d\gamma(h_*)}{dh^2}.
\end{align}
The two expressions are identical. Thus, we conclude that
\begin{align}
	\omega=\gamma_\phi^*-\gamma_{m^2}^*.
\end{align}

Finally, we comment on the relation between Eq.~(\ref{m2eq}) in the superspace formalism and the equation $m^2=m_0^2 \mu^2 Z_\phi Z_{m^2}^{-1}$ mentioned previously for a massive theory. To see the relation between the two, we can rewrite Eq.~(\ref{soft1}) in component fields and compare it with the Lagrangian for a massive complex scalar  
\begin{align}
	\mathcal{L}_{\text{soft}}&=Z|\partial_\mu \phi|^2+Z(m^2-\Delta m^2)|\phi|^2+\dots\nonumber\\
	&=Z_\phi|\partial_\mu \phi|^2+Z_{m^2}m^2|\phi|^2+\dots.
	\label{RelationRG}
\end{align}
In the first line of Eq.~(\ref{RelationRG}) we have inserted the superfield $\tilde{Z}$ into Eq.~(\ref{soft1}) and defined $\Delta m^2\equiv\frac{3}{2}m^2\mu^2h (d\ln Z(h)/dh)$. In the second line, we have written the massive theory in the usual way in terms of a renormalized field $\phi$ and renormalization constants $Z_\phi$ and $Z_{m^2}$. We can thus identify
\begin{align}
	Z_\phi&=Z,\\
	\frac{Z_{m^2}}{Z_\phi}&=1-\frac{\Delta m^2}{m^2} \label{connection}.
\end{align}
The advantage of working in the superspace formalism is that we are able to compute the right-hand side of Eq.~(\ref{connection}) via the relation (\ref{zrelation}) without referring to the mass renormalization constant $Z_{m^2}$ in the massive theory. With these relations, we can easily see that the beta functions (\ref{betam2sf}) and (\ref{betam2_maintext}) are identical.

\section{Coupling to a $U(1)$ gauge field \label{section_gaugefield}}


In this section, we couple the system to a $U(1)$ gauge field. The Lagrangian we consider is
\begin{eqnarray}\label{LGwithA}
\c{L}&=&i\bar{\psi}\slashed{D}\psi+|D_\mu\phi|^2+m^2|\phi|^2+h(\phi^*\psi^Ti\sigma_2\psi+\mathrm{h.c.})\nonumber\\
&&+\lambda^2|\phi|^4+\frac{1}{4}F_{\mu\nu}^2+\frac{1}{2\xi}(\partial_\mu A_\mu)^2,
\end{eqnarray}
where $F_{\mu\nu}=\partial_\mu A_\nu-\partial_\nu A_\mu$ is the field strength tensor, $\xi$ is a gauge-fixing parameter, and we define the gauge-covariant derivatives
\begin{align}\label{DD}
\slashed{D}\psi=(\slashed{\partial}+ie\slashed{A})\psi,
\hspace{5mm}
D_\mu\phi=(\partial_\mu+2ieA_\mu)\phi.
\end{align}
We consider a manifestly Lorentz-invariant Lagrangian, in which the velocities of the fermion ($v_F$), boson ($v_B$), and gauge field ($c$) are all equal and set to one. Even if the bare values of these velocities are different, it was shown in Ref.~\cite{roy2016} that under RG both $v_F$ and $v_B$ flow towards $c$, such that in the deep infrared one ultimately has $v_F=v_B=c=1$. Applying the Halperin-Lubensky-Ma analysis~\cite{halperin1974} to the semimetal-superconductor transition of Dirac fermions, we argue that the coupling to the gauge field destabilizes the SUSY critical point and makes the transition fluctuation-induced first order.

\subsection{Mean-field analysis}
\label{sec:MFgauged}

\begin{figure}
[ptb]
\begin{center}
\includegraphics[width=\columnwidth]{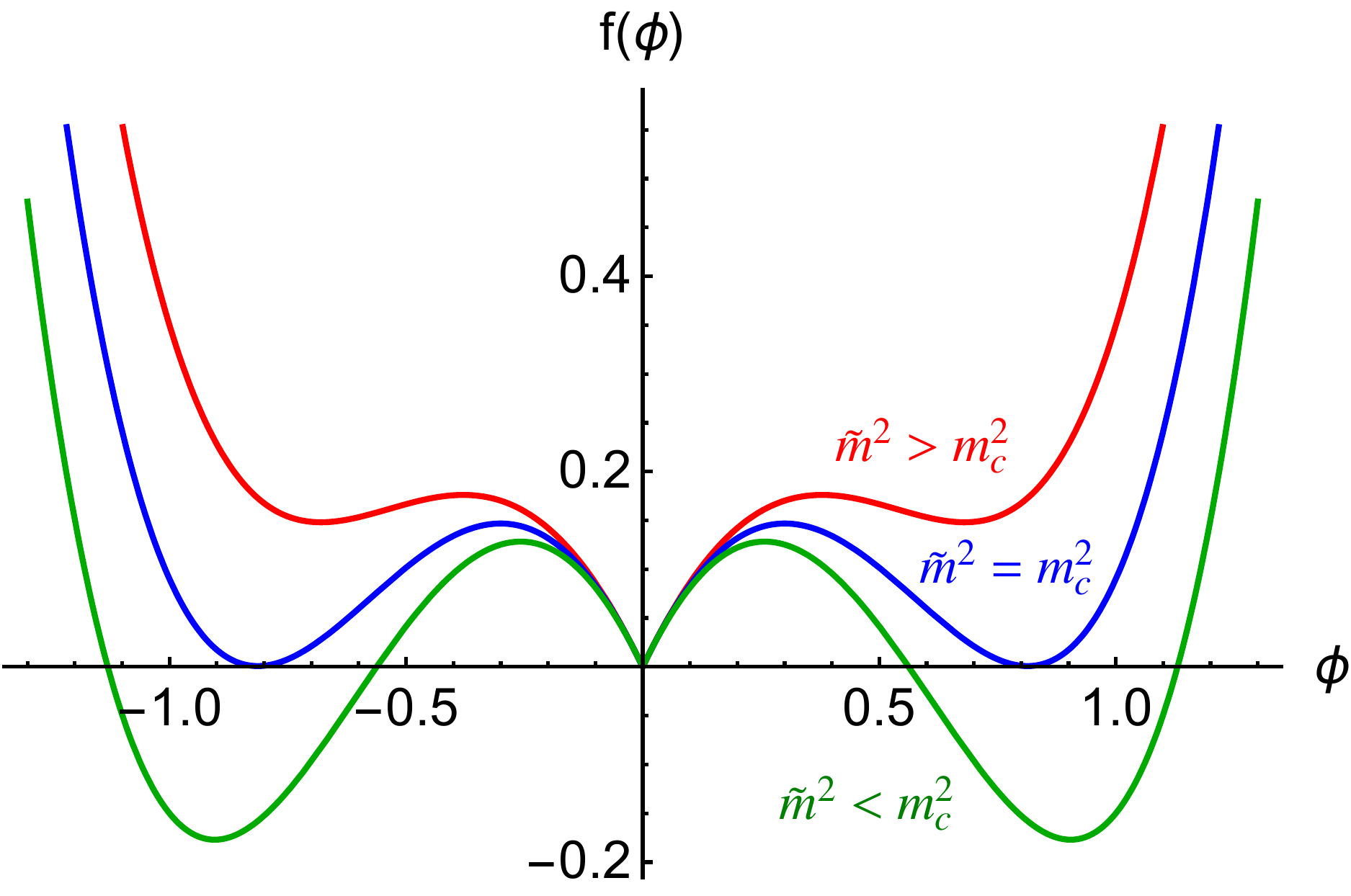}%
\caption{Typical form of the effective Landau free energy $f(\phi)$ obtained by integrating out $U(1)$ gauge fluctuations, where $\tilde{m}^2=m^2+\gamma$ [see Eq.~(\ref{landauFE})]. The cusp at the origin is due to a non-analytic term $\propto|\phi|$ which makes the semimetal-superconductor transition first order [see Eq.~(\ref{landauFE})].}
\label{fig:1storder}
\end{center}
\end{figure}

We first use mean-field arguments to show that the coupling to a dynamical $U(1)$ gauge field destroys the continuous character of the semimetal-superconductor and makes it first order, as in the purely bosonic case. At the mean-field level, we consider that the Cooper pair field $\phi$ is constant and uniform, such that Eq.~(\ref{LGwithA}) becomes
\begin{eqnarray}\label{LGwithA2}
\c{L}&=&i\bar{\psi}\slashed{D}\psi+m^2|\phi|^2+\frac{h}{2}(\phi^*\psi^Ti\sigma_2\psi+\mathrm{h.c.})\nonumber\\
&&+\lambda^2|\phi|^4+\frac{1}{4}F_{\mu\nu}^2+\frac{1}{2\xi}(\partial_\mu A_\mu)^2,
\end{eqnarray}
where we have rescaled $h\rightarrow h/2$ in this section only for convenience. We rewrite the fermionic part of the Lagrangian (\ref{LGwithA2}) in terms of the Nambu spinor $\Psi=(\psi,C\bar{\psi}^T)$, where $C=i\sigma_2$ is a charge-conjugation matrix. Using the property $C\gamma_\mu C=\gamma_\mu^T$, we can write~\cite{kleinert1998}
\begin{align}\label{nambu}
i\bar{\psi}\slashed{D}\psi+\frac{h}{2}(\phi^*\psi^Ti\sigma_2\psi+\mathrm{h.c.})=\frac{1}{2}\Psi^T\mathcal{C}\mathcal{G}^{-1}\Psi,
\end{align}
where $\mathcal{C}=C\oplus C$ and the inverse fermionic propagator is given by $\c{G}^{-1}=\c{G}^{-1}_0+X$ where
\begin{align}
\c{G}_0^{-1}=\begin{pmatrix}
\begin{array}{cc}
h\phi^* & i\slashed{\partial} \\
i\slashed{\partial} & -h\phi
\end{array}
\end{pmatrix},\hspace{5mm}
X=\begin{pmatrix}
\begin{array}{cc}
0 & e\slashed{A} \\
-e\slashed{A} & 0
\end{array}
\end{pmatrix}.
\end{align}
Integrating out $\Psi$ and ignoring a constant contribution arising from the matrix $\c{C}$ in Eq.~(\ref{nambu}), we obtain the bosonic action $S[\phi,A_\mu]=S_0[\phi]+S_A[A_\mu]+S_2[\phi,A_\mu]$ where
\begin{align}
S_0[\phi]&=\int d^3x\left(m^2|\phi|^2+\lambda^2|\phi|^4\right)-\frac{1}{2}\tr\ln\c{G}_0^{-1},
\end{align}
is the bosonic Landau-Ginzburg action in the absence of the gauge field,
\begin{eqnarray}\label{gfaction}
S_A[A_\mu]&=&\frac{1}{2}\int \frac{d^3q}{(2\pi)^3} A_\mu(q)\left[q^2\delta_{\mu\nu}-\left(1-\xi^{-1}\right)q_\mu q_\nu
\right]\nonumber\\
&&\times A_\nu(-q),
\end{eqnarray}
is the action for the gauge field, and
\begin{align}\label{trGXGX}
S_2[\phi,A_\mu]=\frac{1}{4}\tr\c{G}_0X\c{G}_0X+\ldots,
\end{align}
is the one-loop correction to the gauge field action, to quadratic order in the gauge field. In the long-wavelength limit we can evaluate (\ref{trGXGX}) in the gradient expansion. Regulating the ultraviolet divergences in (\ref{trGXGX}) with a momentum cutoff $\Lambda$ and considering that the order parameter is small $h|\phi|\ll\Lambda$, we find to leading order
\begin{align}\label{S'phiA}
S_2[\phi,A_\mu]=\frac{e^2h|\phi|}{4\pi}\int\frac{d^3q}{(2\pi)^3}A_\mu(q)A_\mu(-q),
\end{align}
i.e., the gauge field acquires a mass via the Anderson-Higgs mechanism. We note that by contrast with the purely bosonic problem~\cite{halperin1974}, the gauge field mass is proportional to $|\phi|$ rather than $|\phi|^2$ due to the coupling between $\phi$ and the massless Dirac fermions $\psi$. We now combine Eq.~(\ref{S'phiA}) with Eq.~(\ref{gfaction}) and integrate out the gauge field in the Feynman gauge $\xi=1$. We find the effective Landau free energy $f(\phi)=f_0(\phi)+f_2(\phi)$ where $f_0(\phi)=m^2|\phi|^2+\lambda^2|\phi|^4$ is the free energy in the absence of the gauge field, and $f_2(\phi)$ is the contribution due to gauge field fluctuations, given by
\begin{eqnarray}\label{landauFE}
f_2(\phi)&=&\frac{1}{2}\ln\det\Pi^{-1}(q;\phi)\nonumber\\
&=&\frac{3}{4\pi^2}\int_0^\Lambda dq\,q^2\ln\left(q^2+\frac{e^2h|\phi|}{2\pi}\right)\nonumber\\
&=&\alpha|\phi|-\beta|\phi|^{3/2}+\gamma|\phi|^2+\c{O}(|\phi|^3),
\end{eqnarray}
where $\Pi^{-1}_{\mu\nu}(q;\phi)=(q^2+e^2h|\phi|/2\pi)\delta_{\mu\nu}-(1-\xi^{-1})q_\mu q_\nu$ is the inverse propagator of the gauge field in the presence of the order parameter $\phi$. In Eq.~(\ref{landauFE}) we have assumed $h|\phi|\ll\Lambda$, and the constants $\alpha,\beta,\gamma$ are positive. In addition to the analytic terms $|\phi|^2$, $|\phi|^4$, $\ldots$ already found in the original Lagrangian (\ref{LGwithA2}), we find that non-analytic terms such as $|\phi|$ and $|\phi|^{3/2}$ are generated by gauge field fluctuations. For generic values of $\alpha,\beta,\gamma$, these make the transition first order (Fig.~\ref{fig:1storder}).

\subsection{Renormalization group analysis}
\label{sec:RGgauged}

The mean-field analysis we have just presented neglects the fluctuations of the order parameter $\phi$. To take these fluctuations into account as well as those of the Dirac fermion $\psi$ and gauge field $A_\mu$, we perform a perturbative RG analysis of the gauged Landau-Ginzburg theory (\ref{LGwithA}) on the critical hypersurface $m=0$ in $D=4-\epsilon$ spacetime dimensions.

Following the same procedure as in Sec.~\ref{section_3loop}, we interpret Eq.~(\ref{LGwithA}) as a bare Lagrangian with bare fields and bare couplings denoted by $\psi_0,\phi_0,A_\mu^0$ and $\lambda_0,h_0,e_{0}$, respectively. We denote the corresponding renormalized quantities by $\psi,\phi,A_\mu$ and $\lambda,h,e$. We also denote the bare and renormalized gauge-fixing parameters by $\xi_0$ and $\xi$, respectively. The renormalized Lagrangian is
\begin{eqnarray}\label{gaugedlti2}
\c{L}&=&iZ_\psi\bar{\psi}\slashed{D}\psi+Z_\phi |D_\mu\phi|^2+Z_hh\mu^{\epsilon/2}(\phi^*\psi^Ti\sigma_2\psi+\mathrm{h.c.})\nonumber\\
&&+Z_\lambda^2\lambda^2\mu^\epsilon|\phi|^4+\frac{1}{4}Z_AF_{\mu\nu}^2+\frac{1}{2\xi}(\partial_\mu A_\mu)^2,
\end{eqnarray}
where $Z_A$ is a wave function renormalization constant for the gauge field, and the covariant derivatives are given by
\begin{align}
\slashed{\psi}=(\slashed{\partial}+ie\mu^{\epsilon/2}\slashed{A})\psi,\hspace{2mm}
D_\mu\phi=(\partial_\mu+2ie\mu^{\epsilon/2}A_\mu)\phi.
\end{align}
Gauge invariance (i.e., the Ward identity) implies that there is no separate renormalization constant for the charge $e$ or for the gauge-fixing parameter $\xi$. However, because $Z_\psi,Z_\phi$ depend on the chosen gauge, we have to take into account the $\mu$ dependence of the gauge-fixing parameter $\xi$.

The renormalized Lagrangian can be obtained from the bare one by rescaling the gauge field
\begin{align}
	A_\mu^0&=\sqrt{Z_A}A_\mu,
	\label{coupling2}
\end{align}
and the matter fields as in Eq.~(\ref{wfrgconstant}), which implies the relations (\ref{couplingsh})-(\ref{couplingslambda}) as well as the new relations
\begin{align}
	e^2&=\mu^{-\epsilon}Z_A e_{0}^2, \label{ga} \\
	\xi&=Z_A^{-1}\xi_0. \label{xi} 
\end{align}
Equation (\ref{xi}) is necessary to compute the anomalous dimensions of fields in the gauged theory,
\begin{equation}
	\gamma_x=\beta_{h^2}\frac{\partial\ln Z_x}{\partial h^2}+\beta_{\lambda^2}\frac{\partial\ln Z_x}{\partial\lambda^2}+\beta_\xi \frac{\partial\ln Z_x}{\partial \xi},	
	\label{}
\end{equation}
where $\beta_\xi=d\xi/d\ln\mu$. From the relations (\ref{couplingsh})-(\ref{couplingslambda}) and (\ref{ga}), we can compute the RG beta functions $\beta_{h^2}$, $\beta_{\lambda^2}$, and $\beta_{e^2}$ at the desired order in perturbation theory by inserting the renormalization constants at the same order. 

\begin{figure}
[t]
\begin{center}
\includegraphics[width=0.65\columnwidth]%
{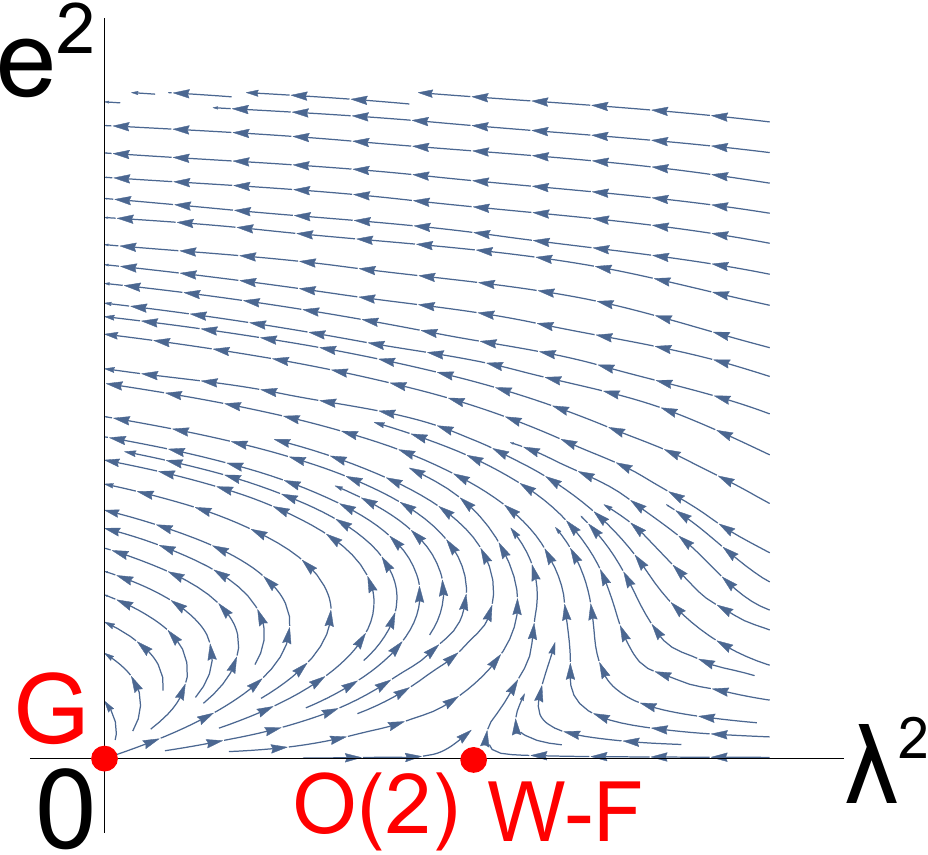}%
\caption{One-loop RG flow for the gauged Landau-Ginzburg theory in the absence of the fermion-boson coupling ($h^2=0)$. The Gaussian and Wilson-Fisher fixed points become unstable when $e^2\neq 0$, giving runaway trajectories towards negative values of the four-boson coupling $\lambda^2$.}
\label{fig:rgflow2}
\end{center}
\end{figure}

\begin{figure}
[t]
\begin{center}
\includegraphics[width=0.65\columnwidth]%
{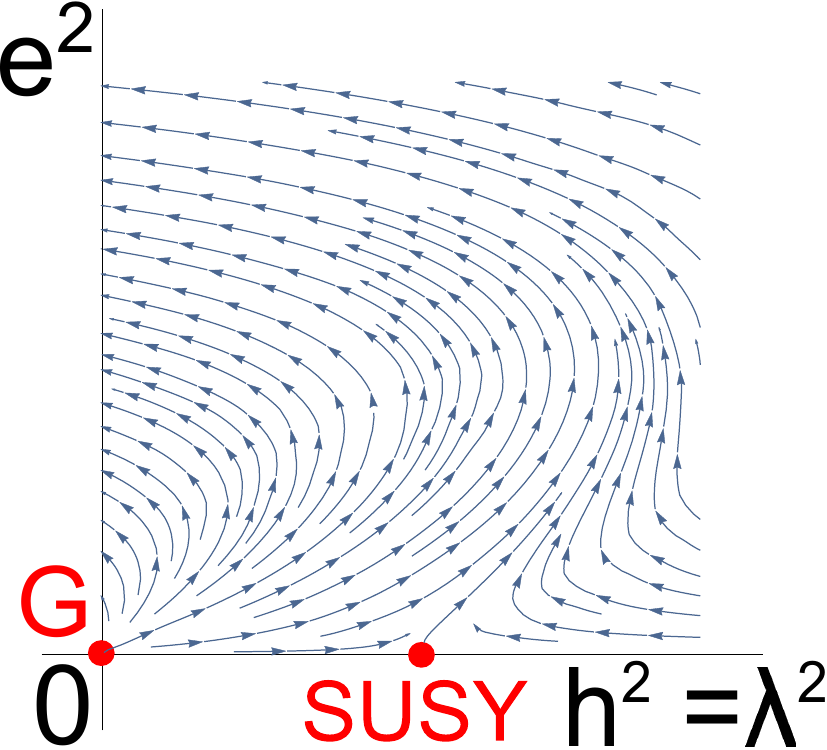}%
\caption{One-loop RG flow for the gauged Landau-Ginzburg theory with finite fermion-boson coupling $h^2$ set equal to the four-boson coupling $\lambda^2$. The SUSY fixed point becomes unstable when $e^2 \neq0$.}
\label{fig:rgflow3}
\end{center}
\end{figure}

At one-loop order, the RG beta functions for the rescaled couplings $\frac{h^2}{(4\pi)^2} \rightarrow h^2$, $\frac{\lambda^2}{(4\pi)^2}\rightarrow\lambda^2$, and $\frac{e^2}{(4\pi)^2}\rightarrow e^2$ are
\begin{align}
	\beta_{h^2}&=-\epsilon h^2-12 e^2 h^2+12 h^4,\label{1loopBetaGaugedh2} \\
	\beta_{\lambda^2}&=-\epsilon \lambda^2 +20\lambda^4 +8h^2 \lambda^2 -16 h^4 -48e^2 \lambda^2+96e^4 ,\label{1loopBetaGaugedl2}\\
	\beta_{e^2}&=-\epsilon e^2 +4e^4,\label{1loopBetaGaugede2}
\end{align}
where we used the one-loop renormalization constants given in Appendix~\ref{app_rgconstant}. On the $e^2=0$ hypersurface, the above beta functions reproduce Eq.~(\ref{1Lbeta1})-(\ref{1Lbeta2}) and the three fixed points found in Sec.~\ref{section_1loop}. For $e^2\neq 0$, all fixed points become imaginary. There is a pair of complex conjugate fixed points with $h_*=0$,
\begin{align}
(h_*^2,\lambda_*^2,e_*^2)=\left(0,\frac{13\pm i\sqrt{311}}{40},\frac{1}{4}\right)\epsilon,
\end{align}
and another pair with $h_*\neq 0$,
\begin{align}
(h_*^2,\lambda_*^2,e_*^2)=\left(\frac{1}{3},\frac{31\pm 3i\sqrt{231}}{120},\frac{1}{4}\right)\epsilon.
\end{align}
These imaginary fixed points correspond to runaway trajectories (Fig.~\ref{fig:rgflow2} and \ref{fig:rgflow3}). In particular, Fig.~\ref{fig:rgflow2} reproduces the known runaway flows in the purely bosonic Abelian Higgs model~\cite{Amit}. As can be seen in Fig.~\ref{fig:rgflow3}, the SUSY fixed point $h_*^2=\lambda_*^2=\frac{\epsilon}{12}$ disappears once the gauge coupling $e^2$ is turned on. Results are qualitatively the same at three-loop order (see Appendix~\ref{app_3loop} for the three-loop RG beta functions of the gauged theory): the Gaussian, Wilson-Fisher, and SUSY fixed points disappear when $e^2\neq 0$ and there are no new real fixed points. As in Ref.~\cite{halperin1974}, we interpret this as the signature of a fluctuation-induced first-order transition.

Of course, one may call into question the validity of both the mean-field theory and the $\epsilon$ expansion. As is well known, the superconducting transition in the (2+1)D bosonic Abelian Higgs model remains continuous in the strongly type-II regime corresponding to $\kappa\gg 1$ where $\kappa\propto\lambda/e$ is the Ginzburg parameter~\cite{kleinert1982}. In this regime, the superconducting transition is dual to the classical 3D XY transition~\cite{dasgupta1981}. The first-order behavior predicted by the $\epsilon$ expansion corresponds to the type-I limit $\kappa\ll 1$. The strongly type-II regime corresponds to the London limit, in which the superconducting coherence length is much shorter than the London penetration depth. Superconducting vortices, which play a key role in duality arguments, contribute significantly to the partition function in this limit, whereas they are suppressed in the opposite (type-I) limit. We thus expect that the conclusion of a fluctuation-induced first-order transition for the gauged Landau-Ginzburg theory (\ref{LGwithA}) with massless Dirac fermions should hold in an appropriate type-I regime for the transition.

Finally, we note that our results disagree with those of Ref.~\cite{roy2013} where a stable, real fixed point with $e_*^2\neq 0$, $h_*^2\neq 0$, and $\lambda_*^2\neq 0$ was found at one-loop order~\footnote{To compare our results with those in Ref. ~\cite{roy2013}, we must make the substitution $(h^2,\lambda^2,e^2)\rightarrow(g^2/8,\lambda/4,e^2/2)$ and use $N_f=1/2$, $N_b=1$. We then obtain Eq.~(10) in Ref.~\cite{roy2013} with $-3e^2g/\epsilon$ on the left-hand side instead of $+3e^2g/\epsilon$, and Eq.~(14) with $-6e^2g^2$ on the right-hand side instead of $-18e^2g^2$.}. In further support of our claims, in Appendix~\ref{app:WilsonRG} we provide an alternative derivation of the one-loop beta functions (\ref{1loopBetaGaugedh2})-(\ref{1loopBetaGaugede2}) using momentum-shell RG.

\subsection{Transition between dual Dirac liquid and T-Pfaffian state}

The results of Sec.~\ref{sec:MFgauged}-\ref{sec:RGgauged} are directly relevant to a subject of much recent interest: correlated surface states of 3D topological insulators and the phase transitions between them. Recently, Ref.~\cite{Wangduality,Metlitskiduality,wang2016,mross2015,karch2016,seiberg2016,murugan2016} proposed a dual description of the 3D topological insulator surface, building on concurrent developments in the theory of the composite Fermi liquid in the half-filled Landau level~\cite{son2015,geraedts2015}. The dual theory is expressed in terms of a single flavor ($N_f=1$) of two-component massless Dirac fermions $\psi_d$ coupled to a noncompact $U(1)$ gauge field $a_\mu$, i.e., QED$_3$,
\begin{eqnarray}
	\mathcal{L}_\text{dual}[\psi_d,a_\mu]&=&\bar{\psi}_d (i\slashed{\partial}+\slashed{a}) \psi_d-\frac{1}{4}(\partial_\mu a_\nu-\partial_\nu a_\mu)^2\nonumber\\
&&	-\frac{1}{4\pi}\epsilon^{\mu\nu\lambda}A_\mu^\text{em}\partial_\nu a_\lambda,
	\label{dualtheory}
\end{eqnarray}
in Minkowski spacetime, where $A_\mu^\text{em}$ is the external (non-dynamical) electromagnetic gauge field, $\bar{\psi}_d=\psi_d^\dagger \gamma^0$ and $\gamma^\mu$ are $2\times 2$ Dirac matrices with $\mu=0,1,2$. Here the dual fermion $\psi_d$ can be viewed as a composite fermion with $4\pi$ flux of the $A_\mu^\text{em}$ gauge field bound to the original electron while a $4\pi$-flux instanton in the dual gauge field $a_\mu$ carries one unit of electron charge $e$ and corresponds to the electron creation operator $\psi^\dag$. Time-reversal symmetry for the original topological insulator is mapped under the duality to charge-conjugation symmetry, such that the dual fermions must be at zero chemical potential. Folllowing Ref.~\cite{Wangduality}, we will refer to the ground state of the dual theory (\ref{dualtheory}) as the dual Dirac liquid. While the exact nature of the dual Dirac liquid is unknown, it is worth mentioning two possibilities: (1) as in the large-$N_f$ limit~\cite{appelquist1986,rantner2001,rantner2002,dipietro2016}, it could be a stable interacting conformal field theory in the infrared~\cite{roscher2016}, i.e., a non-Fermi liquid; (2) it could in fact be dual to the free massless Dirac fermion describing the noninteracting topological insulator surface.

Several known correlated surface states of 3D topological insulators can be accessed from the dual Dirac liquid (\ref{dualtheory}) by adding various interaction terms. In particular, if we add a pairing interaction to condense pairs of $\psi_d$ fermions 
such that $\<\psi_d^T i\sigma_2 \psi_d\> \neq 0$, the resulting surface state in terms of the original electrons is a topologically ordered state called the T-Pfaffian state. This is a gapped phase with non-Abelian topological order that preserves the symmetries of the topological insulator, i.e., $U(1)$ particle-number conservation symmetry and time-reversal symmetry~\cite{bonderson2013,chen2014}. The Landau-Ginzburg theory for the transition between dual Dirac liquid and T-Pfaffian should thus be identical to Eq.~(\ref{LGwithA}), but with $\psi$ replaced by the dual fermion $\psi_d$ and $A_\mu$ replaced by the dual gauge field $a_\mu$. The bosonic order parameter $\phi\sim\langle\psi_d^Ti\sigma_2\psi_d\rangle$ should now be interpreted as the pair amplitude for dual fermions. From our analysis in Sec.~\ref{sec:MFgauged} and \ref{sec:RGgauged}, we conclude that the transition between dual Dirac liquid and T-Pfaffian is fluctuation-induced first order, at least in an analog of the type-I regime for the transition. In this regime the T-Pfaffian would be understood as a type-I Mott insulator~\cite{lee2003} or, more precisely, a $T$-invariant version of the type-I Pfaffian~\cite{parameswaran2011} studied in the context of fractional quantum Hall liquids.

\section{conclusion}
\label{section_conclusion}

In this work we have expanded upon previous studies of the semimetal-superconductor quantum critical point on the surface of a 3D topological insulator in two directions. First, we have established that at three-loop order, the critical dimension below which the SUSY fixed point disappears is $D_c\approx 1.62$, such that at this order in perturbation theory the quantum critical point is indeed endowed with emergent $\mathcal{N}=2$ SUSY. We then derived an exact relation stemming from SUSY between the correlation length and stability critical exponents, which allowed us to obtain the correlation length exponent $\nu$ to order $\epsilon^3$, giving $\nu\approx 0.985$ in 2+1 dimensions. This reduces the difference between the $\epsilon$-expansion result and the conformal bootstrap result $\nu\approx 0.917$ by more than half, compared to the order-$\epsilon$ result $\nu\approx 0.75$. Second, we studied the coupling of the critical Dirac fermion and Cooper pair degrees of freedom to a dynamical $U(1)$ gauge field. By a combination of mean-field and RG arguments, we found that the quantum critical point was destroyed by the gauge fluctuations, most likely replacing it with a fluctuation-induced first-order transition. According to a recent duality conjecture, this gauged Landau-Ginzburg theory was found to also describe the transition between two symmetry-preserving, correlated surface states of 3D topological insulators: the dual Dirac liquid and the T-Pfaffian. Our analysis thus led us to conjecture that the latter transition should also be fluctuation-induced first-order, at least in an appropriate type-I regime for the transition.

\acknowledgements

We thank V. Bouchard, I. F. Herbut, C. Quigley, A. Penin, B. Roy, W. Witczak-Krempa and M. Zoller for useful discussions. This research was supported by NSERC grant \#RGPIN-2014-4608, the Canada Research Chair Program (CRC), the Canadian Institute for Advanced Research (CIFAR), and the University of Alberta.


\appendix

\section{Renormalization constants \label{app_rgconstant}}

We present here the renormalization constants $Z_\psi,Z_\phi,Z_h,Z_\lambda^2$ in the absence of the $U(1)$ gauge field at three-loop order, and the renormalization constants $Z_\psi,Z_\phi,Z_h,Z_\lambda^2,Z_A$ in the presence of the $U(1)$ gauge field at one-loop order only due to the length of the expressions. The full set of renormalization constants at three-loop order for the gauged theory can be made available in electronic format upon request.

In the absence of a gauge field, the renormalization constants at three-loop order are given by
\begin{widetext}
	\begin{eqnarray}
		Z_\psi&=&1-\frac{4 h^2 }{\epsilon } -\frac{16 h^4}{\epsilon ^2}+\frac{8 h^4}{\epsilon } -\frac{320 h^6}{3 \epsilon ^3}+\frac{64h^6}{3 \epsilon ^2}-\left(\frac{\zeta(3)}{2}-\frac{1}{96}\right)\frac{128 h^6 }{\epsilon }+\frac{256 h^4 \lambda ^2}{3 \epsilon ^2}-\frac{128 h^4\lambda ^2}{3 \epsilon }-\frac{32 h^2 \lambda ^4}{3 \epsilon ^2}+\frac{44 h^2 \lambda ^4}{3 \epsilon },\nn\\
		\\
	Z_\phi&=&1-\frac{4 h^2}{\epsilon }-\frac{16 h^4}{\epsilon^2}+\frac{12 h^4}{\epsilon }-\frac{4 \lambda ^4}{\epsilon }-\frac{320h^6}{3 \epsilon ^3}+\frac{112 h^6}{3 \epsilon ^2}-\left(\frac{\zeta(3)}{2}+\frac{5}{96}\right)\frac{128 h^6 }{\epsilon }+\frac{128 h^4 \lambda^2}{\epsilon ^2}-\frac{160 h^4 \lambda ^2}{3 \epsilon }-\frac{16 h^2\lambda ^4}{\epsilon ^2}\nn\\
	&&+\frac{20h^2\lambda^4}{\epsilon}-\frac{160\lambda^6}{3\epsilon^2}+\frac{40\lambda^6}{3\epsilon},\\
	Z_h&=&1
	+\frac{16 h^4}{\epsilon }-\frac{16 h^2 \lambda^2}{\epsilon }
	+\frac{640 h^6}{3 \epsilon^2}-\frac{832 h^6}{3 \epsilon}-\frac{320 h^4 \lambda ^2}{3 \epsilon ^2}+\frac{608 h^4 \lambda^2}{3 \epsilon }-\frac{320 h^2 \lambda ^4}{3 \epsilon ^2}+\frac{224h^2 \lambda^4}{3 \epsilon },\\
	Z_\lambda^2&=& 1
	-\frac{16 h^4}{\epsilon  \lambda ^2}+\frac{20 \lambda^2}{\epsilon }
	-\frac{320 h^4}{\epsilon^2}+\frac{32 h^4}{\epsilon}-\frac{128 h^6}{\epsilon ^2 \lambda ^2}+\frac{128 h^6}{\epsilon\lambda ^2}+\frac{80 h^2 \lambda ^2}{\epsilon ^2}-\frac{40 h^2\lambda ^2}{\epsilon}+\frac{400 \lambda ^4}{\epsilon ^2}-\frac{128 \lambda ^4}{\epsilon }
	-\frac{10240 h^6}{3 \epsilon ^3}\nn\\
	&&+\frac{14336h^6}{3 \epsilon^2}-\left(\frac{68}{3}+8 \zeta(3)\right)\frac{128 h^6 }{\epsilon}+\frac{512 h^8}{\epsilon ^3 \lambda ^2}+\frac{2048 h^8}{3 \epsilon^2 \lambda^2}-\left(1+4 \zeta(3)\right)\frac{256 h^8 }{\epsilon  \lambda ^2}-\frac{8000h^4 \lambda ^2}{\epsilon ^3}+\frac{2528 h^4 \lambda ^2}{\epsilon ^2}\nn\\
	&&+\left(\frac{439}{48}+\frac{17 \zeta(3)}{2}\right)\frac{128h^4  \lambda^2}{\epsilon }+\frac{3200 h^2 \lambda ^4}{\epsilon ^3}-\frac{6848 h^2\lambda ^4}{3\epsilon ^2}+\frac{1024 h^2 \lambda ^4}{3 \epsilon }+\frac{8000\lambda ^6}{\epsilon ^3}-\frac{17600 \lambda ^6}{3 \epsilon^2}+\left(\frac{209}{16}+8\zeta(3)\right)\frac{128  \lambda ^6}{\epsilon }\nn.\\
	\end{eqnarray}
\end{widetext}
Our results can be checked against known results in the limit $\lambda=h$ where SUSY is manifest at the bare level, rather than emergent. In that limit, we find that $Z_h=1$, which is a consequence of the non-renormalization of the superpotential in SUSY theories: see Eq.~(\ref{rigid1}) and the discussion surrounding it. For $\lambda=h$ we also find $Z_\psi=Z_\phi=Z$, in accordance with SUSY. Finally, in this limit our result for $Z$ agrees with Eq.~(3) in Ref.~\cite{avdeev1982} at three-loop order, upon substituting $\epsilon\rightarrow 2\epsilon$ and $h\rightarrow\sqrt{h}/2$ in our result.

In the presence of the gauge field, the renormalization constants at one-loop order are given by
	\begin{align}
		Z_\psi&=1-\frac{4h^2}{\epsilon}-\frac{2e^2 \xi }{\epsilon},\\
		Z_\phi&=1-\frac{4h^2}{\epsilon}+\frac{2e^2(12-4\xi)}{\epsilon},\\
		Z_A&=1-\frac{4e^2}{\epsilon},\\
		Z_h&=1+\frac{6e^2(1-\xi)}{\epsilon},\\
		Z_\lambda^2&=1-\frac{16h^4}{\epsilon \lambda^2}+\frac{20\lambda^2}{\epsilon}-\frac{16e^2 \xi }{\epsilon}+\frac{96e^4}{\epsilon \lambda^2}.
	\end{align}
As a sanity check, if we turn off the gauge coupling by setting $e^2=0$, the two sets of renormalization constants above are the same at one-loop order.

\section{Superspace}\label{app_superspace}

In supersymmetric field theories, various fields are organized into distinct representations of the SUSY algebra called supermultiplets. One can associate to each supermultiplet a superfield which contains all the components of the supermultiplet. In the superspace formalism, the superfield is a function of not only the usual (commuting) spacetime coordinates but also of anticommuting coordinates. In this paper, we consider the theory of a chiral supermultiplet which consists of a complex scalar field $\phi$, a two-component fermionic spinor field $\psi$ and a complex auxiliary scalar field $F$. We can assemble these component fields into a chiral superfield $\Phi$,
\begin{equation}
	\Phi(y)=\phi(y)+\sqrt{2} \theta \psi(y)+\theta^2 F(y),
	\label{superfield}
\end{equation}
where the fields are a function of the superspace coordinate 
\begin{equation}
	y^\mu\equiv x^\mu-i\theta \gamma^\mu\bar{\theta}.
\end{equation}
Here $\theta$ and $\bar{\theta}$ are two-component Grassmann spinors and $\gamma^\mu$ are gamma matrices. For convenience, we define
\begin{align}
	d^2\theta &\equiv -\frac{1}{4}d\theta^\alpha d\theta^\beta \varepsilon_{\alpha\beta}, \\
	d^2\bar{\theta} &\equiv -\frac{1}{4}d\bar{\theta}^\alpha d\bar{\theta}^\beta \varepsilon_{\alpha\beta}, \\
	\theta^2 &\equiv \theta^\alpha \theta_\alpha=\theta^\alpha \varepsilon_{\alpha\beta} \theta^\beta,
\end{align}
where $\varepsilon_{\alpha\beta}=(-i\sigma_2)_{\alpha\beta}$. We then have the Grassmann integrals
\begin{align}
	\int d^2\theta\, \theta^2=1,\  \int d^2\bar{\theta}\, \bar{\theta}^2=1.
	\label{dtheta2}
\end{align}

For $\lambda=h$ and $m^2=0$, the Lagrangian $\mathcal{L}$ in Eq.~(\ref{lti}) is supersymmetric and can be written in the superspace language. First, using the explicit form of the chiral superfield (\ref{superfield}) one can show that
\begin{equation}
	\int d^2\theta d^2\bar{\theta}\,\Phi^\dagger\Phi=i\bar{\psi}\slashed{\partial}\psi+|\partial_{\mu}\phi|^2+ |F|^2,
	\label{}
\end{equation}
modulo a total derivative. To this free theory we add the superpotential $W$, which is a holomorphic function of the chiral superfield,
\begin{equation}
	W(\Phi)=\frac{h}{3}\Phi^3.
	\label{}
\end{equation}
By Taylor expanding in $\theta$ and using Eq.~(\ref{dtheta2}), we obtain
\begin{equation}
\int d^2 \theta\,W(\Phi)=h\phi^2 F+h\phi\psi^Ti\sigma_2\psi,
	\label{}
\end{equation}
modulo a total derivative. Integrating out the auxiliary field $F,F^*$, which is equivalent to using its equation of motion $F=-h\phi^{*2}$, $F^*=-h\phi^2$, we obtain
\begin{equation}
\int d^2\theta d^2\bar{\theta}\,\Phi^\dagger\Phi+\left(\int d^2\theta\,W(\Phi)+\mathrm{h.c.}\right)=\left.\mathcal{L}\right|_{m^2=0,\lambda=h},
\end{equation}
apart from a trivial change of variables $\phi\leftrightarrow\phi^*$.

\section{Three-loop beta functions for the gauged theory}\label{app_3loop}

In this Appendix, we present the three-loop RG beta functions for the gauged Landau-Ginzburg theory (\ref{LGwithA}). The beta functions are given by
\begin{widetext}
	\begin{eqnarray}
	\beta_{e^2}&=&-\epsilon e^2  + 4 e^4 
		+132 e^6-8 e^4 h^2  
		+1404 e^8-716 e^6 h^2+124 e^4 h^4+256 e^6 \lambda ^2-32 e^4\lambda ^4,\\
	\beta_{h^2}&=&-\epsilon h^2 -12 e^2 h^2+12 h^4  
	-234 e^4h^2+372 e^2 h^4+8 h^6-64 h^4 \lambda ^2+8 h^2 \lambda ^4
	+\frac{21014}{3}e^6 h^2-937 e^4 h^4-2104 e^2 h^6\nn\\
	&-&1652 h^8-864\zeta(3) e^6 h^2 -528\zeta(3) e^4 h^4 +576\zeta(3) h^8 -448 e^4 h^2 \lambda^2-1984 e^2 h^4 \lambda^2+1632 h^6 \lambda ^2+320 e^2 h^2 \lambda ^4\nn\\
	&+&300 h^4 \lambda ^4-40h^2 \lambda ^6,\\
	\beta_{\lambda^2}&=&-\epsilon  \lambda ^2
	+96 e^4-16 h^4-48 e^2 \lambda ^2+8 h^2\lambda ^2+20 \lambda ^4
	-4608 e^6+256 e^4 h^2+64e^2 h^4+256h^6+1712 e^4 \lambda ^2+40 e^2 h^2 \lambda ^2\nn\\
	&+&16 h^4 \lambda ^2+448e^2 \lambda ^4-80 h^2 \lambda ^4-240 \lambda ^6
	-74528 e^8+20192e^6 h^2+7304 e^4 h^4+12992 e^2 h^6-768 h^8+76800\zeta(3) e^8 \nn\\
	&+&7680\zeta(3)e^6 h^2 -19968\zeta(3) e^4 h^4 -6144\zeta(3) e^2 h^6 -3072\zeta(3)h^8 +\frac{539336}{3}e^6 \lambda ^2-17894 e^4 h^2 \lambda ^2-16696 e^2 h^4 \lambda^2\nn\\
	&-&8664 h^6 \lambda ^2+18048\zeta(3) e^6  \lambda ^2-1632\zeta(3) e^4 h^2  \lambda^2+18048\zeta(3) e^2 h^4  \lambda ^2-2688\zeta(3) h^6  \lambda^2-64264 e^4 \lambda ^4+1988 e^2 h^2 \lambda ^4\nn\\
	&+&3832 h^4 \lambda^4-36864\zeta(3) e^4 \lambda ^4-3264\zeta(3) e^2 h^2  \lambda ^4+3264\zeta(3) h^4 \lambda ^4-3456 e^2 \lambda ^6+904 h^2 \lambda ^6+4936 \lambda^8+3072 \zeta(3)\lambda ^8.\nn\\
	\end{eqnarray}
\end{widetext}

\section{One-loop momentum-shell renormalization of the gauged theory}
\label{app:WilsonRG}

\fmfset{arrow_len}{2.5mm}
\fmfset{dash_len}{2mm}
\fmfset{wiggly_len}{2mm}
\fmfset{wiggly_slope}{75}

\begin{figure}[t]
\parbox{80pt}{\begin{fmfgraph*}(80,40)
\fmfleft{i}
\fmfright{o}
\fmf{fermion}{i,v1}
\fmf{fermion}{v2,v1}
\fmf{fermion}{v2,o}
\fmf{scalar,left,tension=0}{v1,v2}
\end{fmfgraph*}}
\quad + \quad
\parbox{80pt}{\begin{fmfgraph*}(80,40)
\fmfleft{i1}
\fmfright{o1}
\fmf{fermion}{i1,v1,v2,o1}
\fmf{photon,left,tension=0}{v1,v2}
\end{fmfgraph*}}
\caption{Renormalization of the fermion propagator.}\label{fig:fermion2pt}
\end{figure}
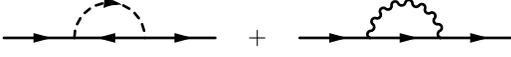

In this Appendix we provide an alternative derivation of the one-loop beta functions (\ref{1loopBetaGaugedh2})-(\ref{1loopBetaGaugede2}) for the gauged Landau-Ginzburg theory (\ref{LGwithA}) using the momentum-shell or Wilsonian RG. In this approach, we first integrate out the high-energy modes $\psi_>,\phi_>,A_\mu^>$ with momentum within the shell $\Lambda/b<|k|<\Lambda$ where $\Lambda$ is the ultraviolet cutoff of the original theory and $b=1+d\ell$ is a positive flow parameter. The contribution $\delta S_<$ to the action for the low-energy modes that is generated by integrating out the high-energy modes is
\begin{eqnarray}
\delta S_<&=&\int d^Dx\biggl(\delta Z_\psi i\bar{\psi}_<\slashed{\partial}\psi_<
+\delta Z_\phi|\partial_\mu\phi_<|^2+\delta Z_\lambda^2\lambda^2|\phi_<|^4 \nonumber\\
&&+\delta Z_hh(\phi^*_<\psi^T_<C\psi_<+\text{h.c.})  +\frac{1}{4}\delta Z_A(F_{\mu\nu}^<)^2 \nonumber\\
&& -\delta Z_eeA_\mu^<\bar{\psi}_<\gamma_\mu\psi_< + \ldots\biggr),\label{LWilsonRG}
\end{eqnarray}
where $D=4-\epsilon$, $C=i\sigma_2$, and $\ldots$ denotes boson-gauge field interactions. We do not write these terms here explicitly as they are not necessary to derive the RG beta functions (\ref{1loopBetaGaugedh2})-(\ref{1loopBetaGaugede2}). The free-field propagators are
\begin{align}
G_{\alpha\beta}(p)&\equiv\langle\psi_\alpha(p)\bar{\psi}_\beta(p)\rangle=\frac{\slashed{p}_{\alpha\beta}}{p^2},\\
D(p)&\equiv\langle\phi(p)\phi^*(p)\rangle=\frac{1}{p^2},\\
\Pi_{\mu\nu}(p)&\equiv\langle A_\mu(p)A_\nu(-p)\rangle=\frac{1}{p^2}\left[\delta_{\mu\nu}+(\xi-1)\frac{p_\mu p_\nu}{p^2}\right].
\end{align}
In Ref.~\cite{roy2013} calculations were performed in the Landau gauge $\xi=0$; here we perform the calculation in a general gauge $\xi$ and show explicitly the gauge invariance of the RG beta functions. In the figures of this section, we draw a single representative diagram to illustrate an entire class of topologically equivalent diagrams. We denote fermion, boson, and gauge field propagators by solid, dashed, and wiggly lines, respectively.

We first renormalize the two-point functions, i.e., the fermion, boson, and gauge field propagators. Two diagrams contribute to the fermion two-point function (Fig.~\ref{fig:fermion2pt}). The first diagram is given by
\begin{align}
\delta Z_\psi^{(1)}\slashed{k}&=4h^2\int_>\frac{d^Dp}{(2\pi)^D}D(p)CG(p+k)^TC\nonumber\\
&=4h^2\left(\frac{D-2}{D}\right)\slashed{k} S_D\Lambda^{-\epsilon}d\ell,
\end{align} 
where $S_D=2/[(4\pi)^{D/2}\Gamma(D/2)]$, we have used the identity $C\gamma_\mu^T C=\gamma_\mu$, and only terms linear in $k$ are kept. The integral is over the momentum shell $\Lambda/b<|p|<\Lambda$. Thus we obtain
\begin{align}\label{dZpsi1}
\delta Z_\psi^{(1)}=2h^2 S_4 d\ell,
\end{align}
to leading order in $\epsilon$. The second diagram in Fig.~\ref{fig:fermion2pt} gives
\begin{align}
\delta Z_\psi^{(2)}\slashed{k}&=-e^2\int_>\frac{d^Dp}{(2\pi)^D}\Pi_{\mu\nu}(p)\gamma_\mu G(p+k)\gamma_\nu\nonumber\\
&=-e^2\left(5-D-\frac{4}{D}-\xi\right)\slashed{k}S_D\Lambda^{-\epsilon}d\ell,
\end{align}
where we used the identity $\gamma_\mu\slashed{p}\gamma_\mu=(2-D)\slashed{p}$. This yields
\begin{align}\label{dZpsi2}
\delta Z_\psi^{(2)}=\xi e^2 S_4 d\ell.
\end{align}

\begin{figure}[t]
\begin{tabular}{c}
\parbox{80pt}{\begin{fmfgraph*}(80,40)
\fmfleft{i}
\fmfright{o}
\fmf{scalar}{i,v,v,o}
\end{fmfgraph*}}
\quad + \quad
\parbox{80pt}{\begin{fmfgraph*}(80,40)
\fmfleft{i}
\fmfright{o}
\fmf{scalar}{i,v1}
\fmf{scalar}{v2,o}
\fmf{fermion,left}{v1,v2}
\fmf{fermion,right}{v1,v2}
\end{fmfgraph*}} \\
+\quad
\parbox{80pt}{\begin{fmfgraph*}(80,40)
\fmfleft{i1}
\fmfright{o1}
\fmf{scalar}{i1,v1,v2,o1}
\fmf{photon,left,tension=0}{v1,v2}
\end{fmfgraph*}}
\quad + \quad
\parbox{80pt}{\begin{fmfgraph*}(80,40)
\fmfleft{i}
\fmfright{o}
\fmf{scalar}{i,v}
\fmf{scalar}{v,o}
\fmf{photon}{v,v}
\end{fmfgraph*}}
\end{tabular}
\caption{Renormalization of the boson propagator.}\label{fig:boson2pt}
\end{figure}
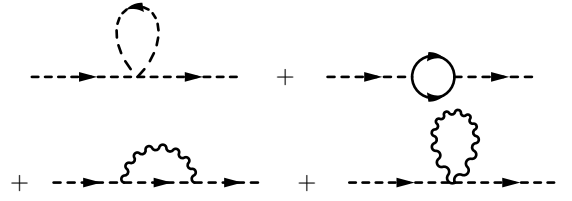

Four diagrams contribute to the boson two-point function (Fig.~\ref{fig:boson2pt}). The first and fourth (tadpole) diagrams in Fig.~\ref{fig:boson2pt} only contribute to the boson mass $m^2$, which we neglect here as we are interested in the critical theory. The second diagram gives
\begin{align}
\delta Z_\phi^{(2)}k^2&=2h^2\int_>\frac{d^Dp}{(2\pi)^D}\tr G(p)CG(-p+k)^TC\nonumber\\
&=4h^2\left(\frac{D-2}{D}\right)k^2S_D\Lambda^{-\epsilon}d\ell,
\end{align}
hence
\begin{align}
\delta Z_\phi^{(2)}=2h^2 S_4d\ell.
\end{align}
The third diagram gives
\begin{align}
\delta Z_\phi^{(3)}k^2&=-4e^2\int_>\frac{d^Dp}{(2\pi)^D}D(p-k)\Pi_{\mu\nu}(p)\nonumber\\
&\hspace{10mm}\times[p_\mu p_\nu-2(p_\mu k_\nu+p_\nu k_\mu)+4k_\mu k_\nu]\nonumber\\
&=-4e^2\left(4-\frac{4}{D}-\xi\right)k^2S_D\Lambda^{-\epsilon}d\ell,
\end{align}
and we obtain
\begin{align}
\delta Z_\phi^{(3)}=-4e^2(3-\xi)S_4d\ell.
\end{align}

\begin{figure}[t]
\begin{tabular}{c}
\parbox{80pt}{\begin{fmfgraph*}(80,40)
\fmfleft{i}
\fmfright{o}
\fmf{photon}{i,v1}
\fmf{photon}{v2,o}
\fmf{fermion,left}{v1,v2}
\fmf{fermion,left}{v2,v1}
\end{fmfgraph*}}
\quad + \quad
\parbox{80pt}{\begin{fmfgraph*}(80,40)
\fmfleft{i}
\fmfright{o}
\fmf{photon}{i,v1}
\fmf{photon}{v2,o}
\fmf{scalar,left}{v1,v2}
\fmf{scalar,left}{v2,v1}
\end{fmfgraph*}} \\
+\quad
\parbox{80pt}{\begin{fmfgraph*}(80,40)
\fmfleft{i}
\fmfright{o}
\fmf{photon}{i,v}
\fmf{photon}{v,o}
\fmf{scalar}{v,v}
\end{fmfgraph*}}
\end{tabular}
\caption{Renormalization of the gauge field propagator.}\label{fig:photon2t}
\end{figure}
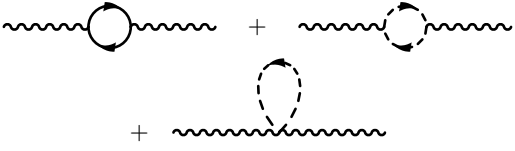

Finally, the renormalization of the gauge field propagator is given by the diagrams in Fig.~(\ref{fig:photon2t}). The first diagram gives
\begin{align}\label{MomShellGaugeInv}
\delta &Z_A^{(1)}(q^2\delta_{\mu\nu}-q_\mu q_\nu)=e^2\int_>\frac{d^Dp}{(2\pi)^D}\tr\gamma_\mu G(p)\gamma_\nu G(p+q)\nonumber\\
&=2\left(\frac{D-2}{D+2}\right)\left(q^2\delta_{\mu\nu}-\frac{4}{D}q_\mu q_\nu\right)S_D\Lambda^{-\epsilon}d\ell,
\end{align}
where only terms quadratic in $q$ are kept. In Eq.~(\ref{MomShellGaugeInv}), quadratically divergent integrals $\sim\Lambda^2$ were evaluated in $D=2$, in the sense that
\begin{align}
\int\frac{d^Dp}{(2\pi)^D}\frac{p_\mu p_\nu}{(p^2)^2}=\frac{1}{2}\delta_{\mu\nu}\int\frac{d^Dp}{(2\pi)^D}\frac{1}{p^2},
\end{align}
where the right-hand side contains $\delta_{\mu\nu}/2$ rather than $\delta_{\mu\nu}/D$ with $D\rightarrow 4$. This was shown to be a consistent procedure for restoring gauge invariance in theories with a hard momentum cutoff $\Lambda$~\cite{varin2007}. Alternatively, one can simply discard the non-gauge invariant mass term $\sim \Lambda^2A_\mu^2$ that is generated by momentum-shell integration~\cite{halperin1974,chen1978}. Setting $D=4$ in Eq.~(\ref{MomShellGaugeInv}), the correction to the gauge field propagator is purely transverse, giving
\begin{align}
\delta Z_A^{(1)}=\frac{2e^2}{3}S_4d\ell.
\end{align}
There is no renormalization $\delta\xi$ of the gauge-fixing parameter, as in field-theoretic RG~\cite{Justinbook}. The second diagram in Fig.~\ref{fig:photon2t} is given by
\begin{align}\label{dZA2}
&\delta Z_A^{(2)}(q^2\delta_{\mu\nu}-q_\mu q_\nu)\nonumber\\
&\hspace{5mm}=-8e^2\int_>\frac{d^Dp}{(2\pi)^D}D(p)D(p+q)p_\mu(2p+q)_\nu\nonumber\\
&\hspace{5mm}=4e^2S_4d\ell\left[-\Lambda^2\delta_{\mu\nu}+\frac{1}{3}(q^2\delta_{\mu\nu}-q_\mu q_\nu)\right],
\end{align}
in $D=4$. The gauge field mass term $\sim\Lambda^2\delta_{\mu\nu}$ explicitly violates gauge invariance. The third diagram in Fig.~(\ref{fig:photon2t}) is given by
\begin{align}
\delta Z_A^{(3)}(q^2\delta_{\mu\nu}-q_\mu q_\nu)=4e^2S_4d\ell\Lambda^2\delta_{\mu\nu},
\end{align}
which also violates gauge invariance. However, adding the two diagrams restores gauge invariance,
\begin{align}
\delta Z_A^{(2)}+\delta Z_A^{(3)}=\frac{4e^2}{3}S_4d\ell.
\end{align}

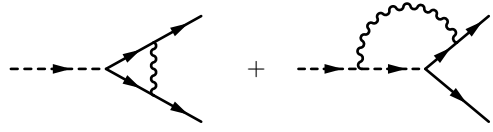
\begin{figure}[t]
\parbox{80pt}{\begin{fmfgraph*}(80,40)
\fmfleft{i}
\fmfright{o1,o2}
\fmf{scalar}{i,v}
\fmf{fermion}{v,v1}
\fmf{fermion}{v1,o1}
\fmf{fermion}{v,v2}
\fmf{fermion}{v2,o2}
\fmf{photon,tension=0}{v1,v2}
\end{fmfgraph*}}\quad + \quad
\parbox{80pt}{\begin{fmfgraph*}(80,40)
\fmfleft{i}
\fmfright{o1,o2}
\fmf{scalar}{i,v1}
\fmf{scalar}{v1,v}
\fmf{fermion}{v,v2}
\fmf{fermion}{v2,o2}
\fmf{fermion,tension=1/2}{v,o1}
\fmf{photon,left,tension=0}{v1,v2}
\end{fmfgraph*}}
\caption{Renormalization of the fermion-boson vertex.}\label{fig:fb3pt}
\end{figure}

\begin{figure}[t]
\parbox{80pt}{\begin{fmfgraph*}(80,40)
\fmfleft{i}
\fmfright{o1,o2}
\fmf{photon}{i,v}
\fmf{fermion}{v,v1}
\fmf{fermion}{v1,o1}
\fmf{fermion}{v2,v}
\fmf{fermion}{o2,v2}
\fmf{photon,tension=0}{v1,v2}
\end{fmfgraph*}}\quad + \quad
\parbox{80pt}{\begin{fmfgraph*}(80,40)
\fmfleft{i}
\fmfright{o1,o2}
\fmf{photon}{i,v}
\fmf{fermion}{v1,v}
\fmf{fermion}{v1,o1}
\fmf{fermion}{v,v2}
\fmf{fermion}{o2,v2}
\fmf{scalar,tension=0}{v2,v1}
\end{fmfgraph*}} 
\quad + \quad
\parbox{80pt}{\begin{fmfgraph*}(80,40)
\fmfleft{i}
\fmfright{o1,o2}
\fmf{photon}{i,v}
\fmf{scalar}{v,v1}
\fmf{fermion}{v1,o1}
\fmf{scalar}{v2,v}
\fmf{fermion}{o2,v2}
\fmf{fermion,tension=0}{v1,v2}
\end{fmfgraph*}}
\caption{Renormalization of the fermion-gauge field vertex.}\label{fig:fgf3pt}
\end{figure}
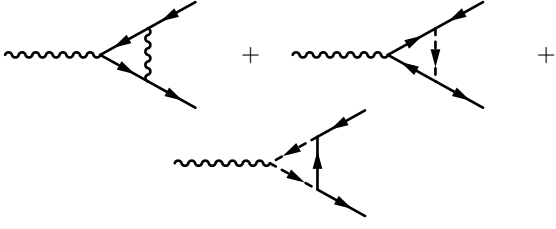

We now turn to the renormalization of the three-point functions: the fermion-boson vertex (Fig.~\ref{fig:fb3pt}) and the fermion-gauge field vertex (Fig.~\ref{fig:fgf3pt}). There is no renormalization of the fermion-boson vertex at one loop in the ungauged theory: the analog of the first diagram in Fig.~\ref{fig:fb3pt} with the gauge field propagator replaced by a boson propagator is incompatible with the Feynman rules of the theory, i.e., the direction of the fermionic flow at the bare fermion-boson vertex. The first diagram in Fig.~\ref{fig:fb3pt} is given by
\begin{align}
\delta Z_h^{(1)}h\mathbb{I}&=he^2\int_>\frac{d^Dp}{(2\pi)^D}\Pi_{\mu\nu}(p)\gamma_\mu G(p)G(-p)\gamma_\nu\nonumber\\
&=-(D+\xi-1)he^2\mathbb{I}S_D\Lambda^{-\epsilon}d\ell,
\end{align}
where $\mathbb{I}$ denotes the unit $2\times 2$ matrix appearing in the Clifford algebra $\{\gamma_\mu,\gamma_\nu\}=2\delta_{\mu\nu}\mathbb{I}$. This yields
\begin{align}
\delta Z_h^{(1)}=-(3+\xi)e^2 S_4 d\ell.
\end{align}
The second diagram in Fig.~\ref{fig:fb3pt} gives
\begin{align}
\delta Z_h^{(2)}h\mathbb{I}&=4he^2\int_>\frac{d^Dp}{(2\pi)^D}D(p)\Pi_{\mu\nu}(p)p_\nu G(p)\gamma_\mu\nonumber\\
&=4\xi he^2\mathbb{I}S_D\Lambda^{-\epsilon}d\ell,
\end{align}
hence we obtain
\begin{align}
\delta Z_h^{(2)}=4\xi e^2 S_4 d\ell.
\end{align}

The renormalization of the fermion-gauge field vertex (Fig.~\ref{fig:fgf3pt}) will allow us to verify the Ward identity $\delta Z_e=\delta Z_\psi$~\cite{PeskinSchroeder}, which is a consequence of the $U(1)$ gauge invariance of the theory. The first diagram in Fig.~\ref{fig:fgf3pt} is given by
\begin{align}
\delta Z_e^{(1)}e\gamma_\mu&=e^3\int_>\frac{d^Dp}{(2\pi)^D}\Pi_{\nu\lambda}(p)\gamma_\nu G(p)\gamma_\mu G(p)\gamma_\lambda\nn\\
&=-e^3\left(5-D-\frac{4}{D}-\xi\right)\gamma_\mu S_D\Lambda^{-\epsilon}d\ell,
\end{align}
where we used the identity $\slashed{q}\gamma_\mu\slashed{q}=2q_\mu\slashed{q}-q^2\gamma_\mu$. Thus,
\begin{align}
\delta Z_e^{(1)}=\xi e^2S_4d\ell.
\end{align}
Comparing with Eq.~(\ref{dZpsi2}), we see that $\delta Z_e^{(1)}=\delta Z_\psi^{(2)}$, which is the well-known Ward identity in QED at one-loop order~\cite{PeskinSchroeder} in the absence of a fermion-boson coupling. However, the Ward identity is expected to hold even in the presence of the fermion-boson coupling, as the latter preserves the gauge invariance of the theory. We thus expect that the first diagram in Fig.~\ref{fig:fermion2pt} should be equal to the sum of the last two diagrams in Fig.~\ref{fig:fgf3pt}. The second diagram in Fig.~\ref{fig:fgf3pt} is given by
\begin{align}
\delta Z_e^{(2)}e\gamma_\mu&=4eh^2\int_>\frac{d^Dp}{(2\pi)^D}D(p)CG^T(p)\gamma_\mu^TG^T(p)C\nn\\
&=4eh^2\left(\frac{2-D}{D}\right)\gamma_\mu S_D\Lambda^{-\epsilon}d\ell.
\end{align}
As a result,
\begin{align}
\delta Z_e^{(2)}=-2h^2S_4d\ell.
\end{align}
The third diagram in Fig.~\ref{fig:fgf3pt} is given by
\begin{align}
\delta Z_e^{(3)}e\gamma_\mu&=16eh^2\int_>\frac{d^Dp}{(2\pi)^D}D(p)^2p_\mu CG^T(p)C\nn\\
&=\frac{16}{D}eh^2\gamma_\mu S_D\Lambda^{-\epsilon}d\ell,
\end{align}
thus
\begin{align}
\delta Z_e^{(3)}=4h^2S_4d\ell.
\end{align}
Comparing with Eq.~(\ref{dZpsi1}), we find $\delta Z_e^{(2)}+\delta Z_e^{(3)}=\delta Z_\psi^{(1)}$, as expected.

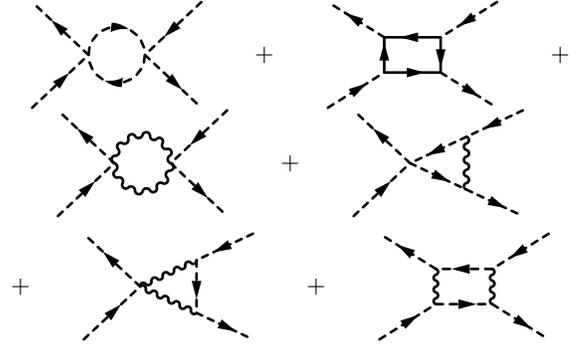
\begin{figure}[t]
\parbox{80pt}{\begin{fmfgraph*}(80,40)
\fmfleft{i1,i2}
\fmfright{o1,o2}
\fmf{scalar}{i1,v1}
\fmf{scalar}{v1,i2}
\fmf{scalar,left}{v1,v2}
\fmf{scalar,left}{v2,v1}
\fmf{scalar}{v2,o1}
\fmf{scalar}{o2,v2}
\end{fmfgraph*}}
\quad + \quad
\parbox{80pt}{\begin{fmfgraph*}(80,40)
\fmfleft{i1,i2}
\fmfright{o1,o2}
\fmf{scalar}{i1,v1}
\fmf{scalar}{v2,i2}
\fmf{scalar}{v3,o1}
\fmf{scalar}{o2,v4}
\fmf{fermion}{v1,v2}
\fmf{fermion}{v4,v2}
\fmf{fermion}{v4,v3}
\fmf{fermion}{v1,v3}
\end{fmfgraph*}} 
\quad + \quad
\parbox{80pt}{\begin{fmfgraph*}(80,40)
\fmfleft{i1,i2}
\fmfright{o1,o2}
\fmf{scalar}{i1,v1}
\fmf{scalar}{v1,i2}
\fmf{photon,left}{v1,v2}
\fmf{photon,right}{v1,v2}
\fmf{scalar}{v2,o1}
\fmf{scalar}{o2,v2}
\end{fmfgraph*}}
\quad + \quad
\parbox{80pt}{\begin{fmfgraph*}(80,40)
\fmfleft{i1,i2}
\fmfright{o1,o2}
\fmf{scalar}{i1,v1}
\fmf{scalar}{v1,i2}
\fmf{scalar}{v1,v2}
\fmf{scalar}{v3,v1}
\fmf{photon,tension=0}{v2,v3}
\fmf{scalar}{v2,o1}
\fmf{scalar}{o2,v3}
\end{fmfgraph*}} \\
\vspace{2mm}
 + \quad
\parbox{80pt}{\begin{fmfgraph*}(80,40)
\fmfleft{i1,i2}
\fmfright{o1,o2}
\fmf{scalar}{i1,v1}
\fmf{scalar}{v1,i2}
\fmf{photon}{v1,v2}
\fmf{photon}{v1,v3}
\fmf{scalar}{v2,o1}
\fmf{scalar}{o2,v3}
\fmf{scalar,tension=0}{v3,v2}
\end{fmfgraph*}}
\quad + \quad
\parbox{80pt}{\begin{fmfgraph*}(80,40)
\fmfleft{i1,i2}
\fmfright{o1,o2}
\fmf{scalar}{i1,v1}
\fmf{scalar}{v2,i2}
\fmf{scalar}{v3,o1}
\fmf{scalar}{o2,v4}
\fmf{photon}{v1,v2}
\fmf{photon}{v3,v4}
\fmf{scalar}{v1,v3}
\fmf{scalar}{v4,v2}
\end{fmfgraph*}}
\caption{Renormalization of the four-boson vertex.}\label{fig:boson4pt}
\end{figure}

We finally turn to the renormalization of the boson four-point function (Fig.~\ref{fig:boson4pt}). The first diagram is standard from the $O(2)$ vector model~\cite{Sachdev}, and we have
\begin{align}
\delta Z_\lambda^{2(1)}\lambda^2&=-10\lambda^4\int_>\frac{d^Dp}{(2\pi)^D} D(p)^2\nn\\
&=-10\lambda^4 S_D\Lambda^{-\epsilon}d\ell,
\end{align}
thus we obtain
\begin{align}
\delta Z_\lambda^{2(1)}=-10\lambda^2 S_4d\ell.
\end{align}
The second diagram in Fig.~\ref{fig:boson4pt} is
\begin{align}
&\delta Z_\lambda^{2(2)}\lambda^2\nn\\
&\hspace{5mm}=4h^4\int_>\frac{d^Dp}{(2\pi)^D}\tr G(p)CG(-p)^TCG(p)CG(-p)^TC\nn\\
&\hspace{5mm}=8h^4 S_D\Lambda^{-\epsilon}d\ell,
\end{align}
giving
\begin{align}
\delta Z_\lambda^{2(2)}=8h^4\lambda^{-2}S_4d\ell,
\end{align}
where the negative power of $\lambda^2$ simply indicates that a four-boson coupling can be generated by a fermion loop, at zeroth order in $\lambda^2$. The third diagram in Fig.~\ref{fig:boson4pt} is given by
\begin{align}
\delta Z_\lambda^{2(3)}\lambda^2&=-16e^4\int_>\frac{d^Dp}{(2\pi)^D}\Pi_{\mu\nu}(p)\Pi_{\mu\nu}(p)\nn\\
&=-16e^4(D-1+\xi^2)S_D\Lambda^{-\epsilon}d\ell,
\end{align}
hence
\begin{align}
\delta Z_\lambda^{2(3)}=-16e^4\lambda^{-2}(3+\xi^2)S_4d\ell.
\end{align}
Likewise, here the negative power of $\lambda^2$ indicates that a four-boson coupling can be generated by a gauge field loop. The fourth diagram in Fig.~\ref{fig:fgf3pt} is given by
\begin{align}
\delta Z_\lambda^{2(4)}\lambda^2&=8e^2\lambda^2\int_>\frac{d^Dp}{(2\pi)^D}D(p)^2 p_\mu p_\nu\Pi_{\mu\nu}(p)\nn\\
&=8e^2\lambda^2\xi S_D\Lambda^{-\epsilon}d\ell,
\end{align}
which implies
\begin{align}
\delta Z_\lambda^{2(4)}=8\xi e^2S_4d\ell.
\end{align}
The fifth diagram in Fig.~\ref{fig:boson4pt} is given by
\begin{align}
\delta Z_\lambda^{2(5)}\lambda^2&=32e^4\int_>\frac{d^Dp}{(2\pi)^D}D(p)p_\nu p_\lambda\Pi_{\mu\nu}(p)\Pi_{\mu\lambda}(p)\nn\\
&=32e^4\xi^2 S_D\Lambda^{-\epsilon}d\ell,
\end{align}
thus
\begin{align}
\delta Z_\lambda^{2(5)}=32\xi^2e^4\lambda^{-2}S_4d\ell.
\end{align}
Finally, the sixth diagram in Fig.~\ref{fig:boson4pt} is given by
\begin{align}
\delta Z_\lambda^{2(6)}\lambda^2&=-16e^4\int_>\frac{d^Dp}{(2\pi)^D}D(p)^2 p_\mu p_\nu p_\lambda p_\rho\Pi_{\mu\nu}(p)\Pi_{\lambda\rho}(p)\nn\\
&=-16e^4\xi^2 S_D\Lambda^{-\epsilon}d\ell,
\end{align}
and we obtain
\begin{align}
\delta Z_\lambda^{2(6)}=-16\xi^2e^4\lambda^{-2}S_4d\ell.
\end{align}

Adding the contributions from all diagrams, the renormalization constants are given by
\begin{align}
\delta Z_\psi&=(2h^2+\xi e^2)S_4d\ell,\label{RC1}\\
\delta Z_\phi&=(2h^2-12e^2+4\xi e^2)S_4d\ell,\\
\delta Z_A&=2e^2S_4d\ell,\\
\delta Z_h&=(-3e^2+3\xi e^2)S_4d\ell,\\
\delta Z_e&=(2h^2+\xi e^2)S_4d\ell,\\
\delta Z_\lambda^2&=\left(-10\lambda^2+8h^4\lambda^{-2}-48 e^4\lambda^{-2}+8\xi e^2\right)S_4d\ell.\label{RC6}
\end{align}
As observed before, the Ward identity $\delta Z_\psi=\delta Z_e$ is satisfied. The non-gauge invariant terms $\propto\xi^2$ disappear from $\delta Z_\lambda^2$, as a result of cancellations between the three diagrams in Fig.~\ref{fig:boson4pt} with two internal gauge field propagators. The infrared RG beta functions are given in terms of the renormalization constants by~\cite{Sachdev}
\begin{align}
\frac{dh^2}{d\ell}&=\epsilon h^2+h^2\left(2\frac{\delta Z_h}{d\ell}-2\frac{\delta Z_\psi}{d\ell}-\frac{\delta Z_\phi}{d\ell}\right),\\
\frac{d\lambda^2}{d\ell}&=\epsilon\lambda^2+\lambda^2\left(\frac{\delta Z_\lambda^2}{d\ell}-2\frac{\delta Z_\phi}{d\ell}\right),\\
\frac{de^2}{d\ell}&=\epsilon e^2+e^2\left(2\frac{\delta Z_e}{d\ell}-2\frac{\delta Z_\psi}{d\ell}-\frac{\delta Z_A}{d\ell}\right).
\end{align}
Substituting in the renormalization constants (\ref{RC1})-(\ref{RC6}) and defining the rescaled couplings $h^2\rightarrow h^2/(4\pi)^2$, $\lambda\rightarrow\lambda/(4\pi)^2$, and $e^2\rightarrow e^2/(4\pi)^2$, we obtain
\begin{align}
\frac{dh^2}{d\ell}&=\epsilon h^2+12e^2h^2-12h^4,\label{betah2Wilson}\\
\frac{d\lambda^2}{d\ell}&=\epsilon\lambda^2-20\lambda^4-8h^2\lambda^2+16h^4+48e^2\lambda^2-96e^4,\label{betalambda2Wilson}\\
\frac{de^2}{d\ell}&=\epsilon e^2-4e^4,\label{betae2Wilson}
\end{align}
where we used $S_4=2/(4\pi)^2$. All the non-gauge invariant $\xi$-dependent terms drop out of the beta functions. Using $d\ell=-d\ln\mu$, one precisely recovers the beta functions in Eq.~(\ref{1loopBetaGaugedh2})-(\ref{1loopBetaGaugede2}). Equations~(\ref{betalambda2Wilson}) and (\ref{betae2Wilson}) agree with Eq.~(13) and (15) in Ref.~\cite{roy2013}, but Eq.~(\ref{betah2Wilson}) disagrees with the corresponding Eq.~(14) in that paper.

\section{Calculation of the renormalization constants at three-loop order: technical aspects}
\label{app_rg}

\renewcommand{\picscalefactor}{0.15}
\begin{table}[t]
\centering
\begin{tabular}{c|c|c|c}
Loops & 1 & 2 & 3\\
\hline
 \vcenteredhbox{\includegraphics[scale=\picscalefactor]{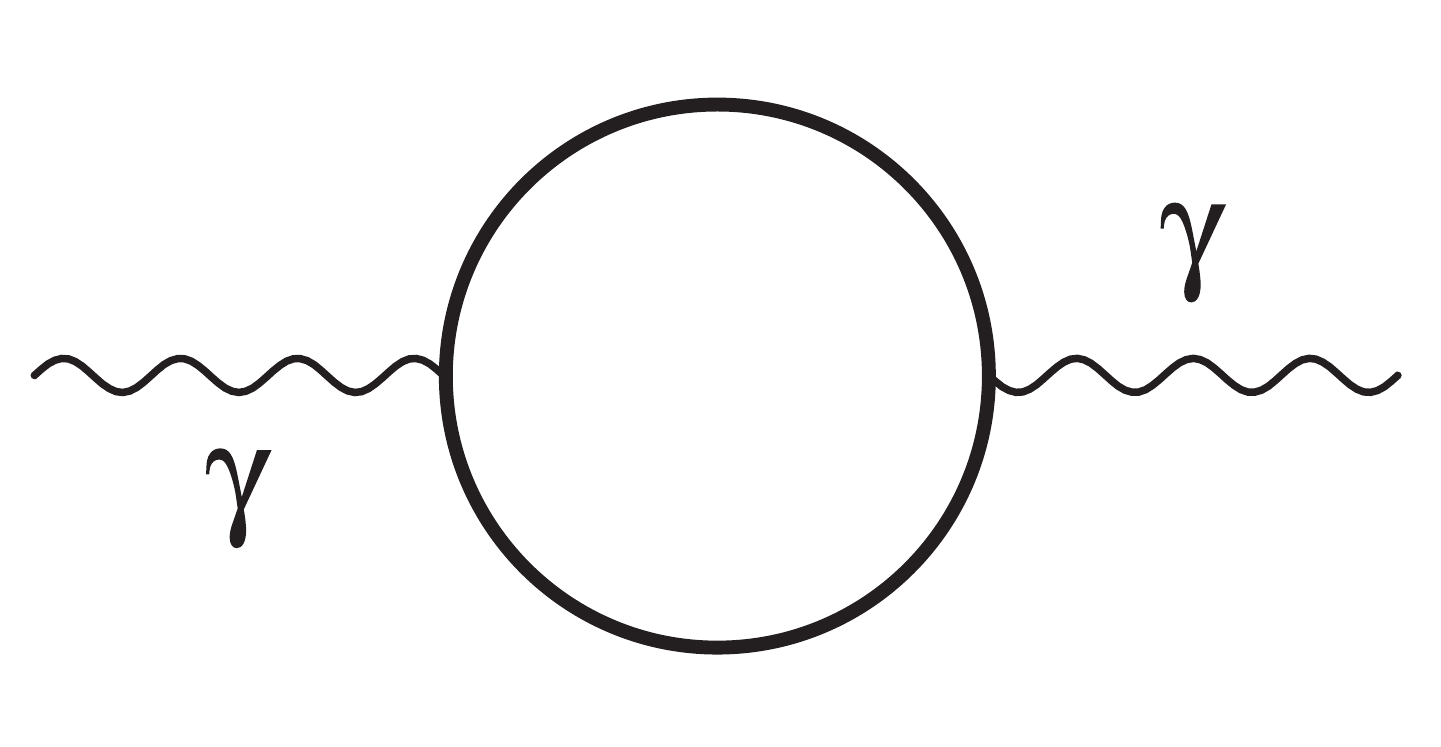}} & \vcenteredhbox{$3$} & \vcenteredhbox{$27$} & \vcenteredhbox{$502$}\\
\hline
 \vcenteredhbox{\includegraphics[scale=\picscalefactor]{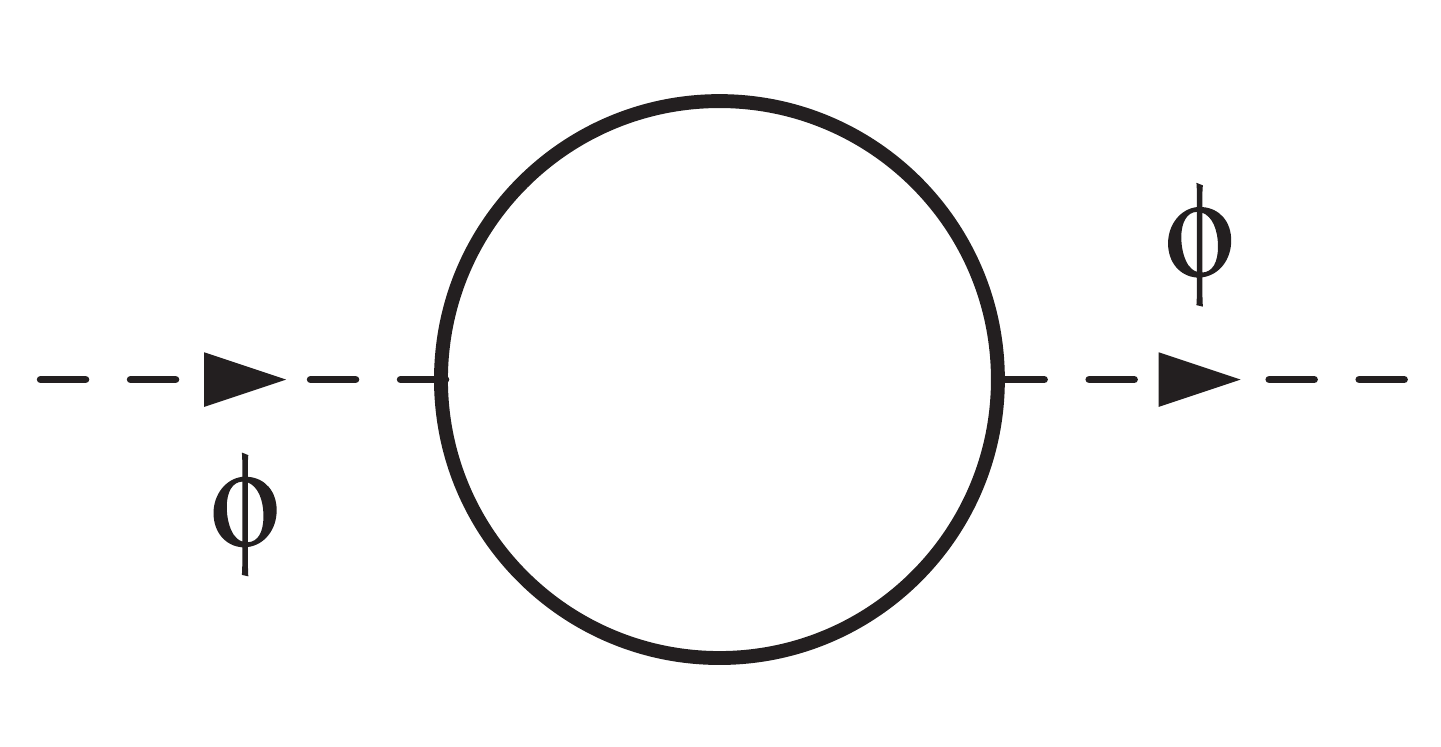}}  & \vcenteredhbox{$4$} & \vcenteredhbox{$27$} & \vcenteredhbox{$455$}\\
\hline
 \vcenteredhbox{\includegraphics[scale=\picscalefactor]{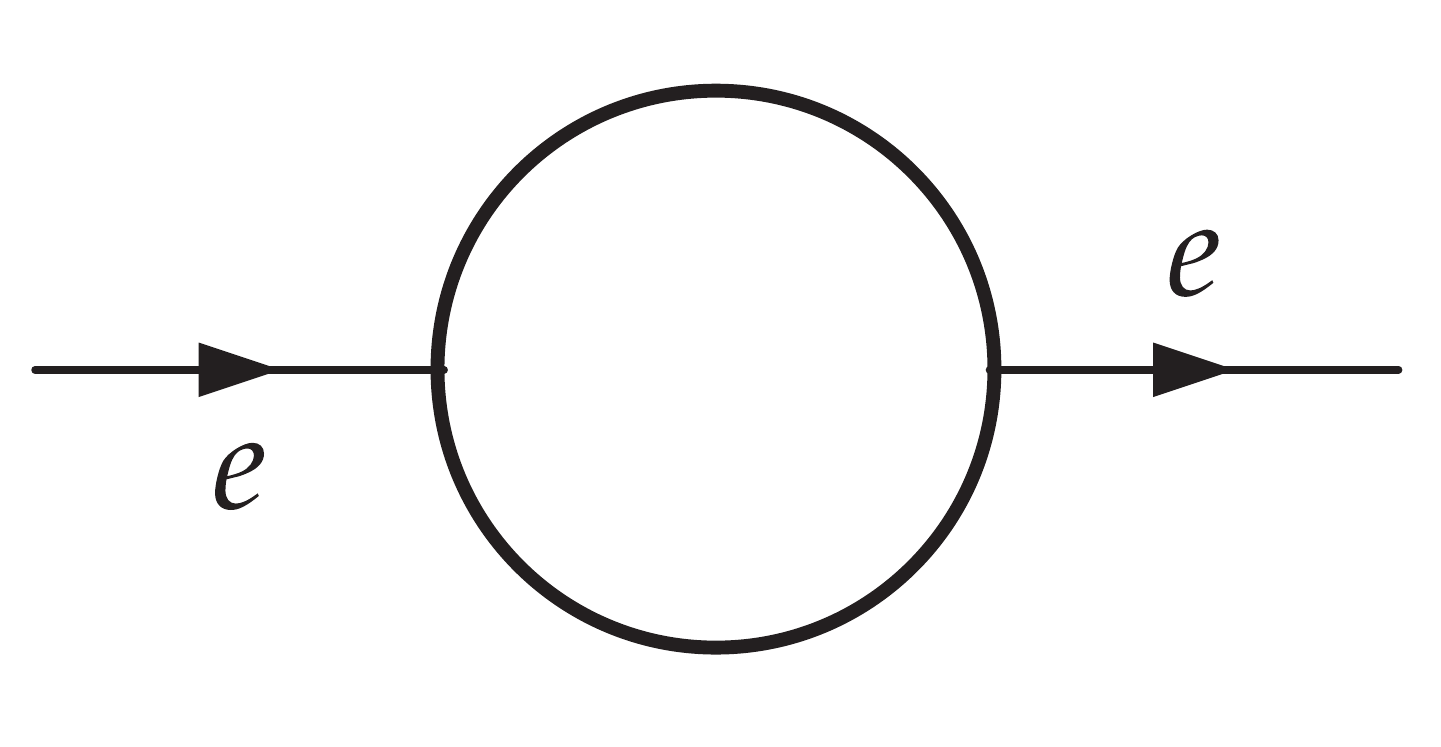}} & \vcenteredhbox{$2$} & \vcenteredhbox{$17$} & \vcenteredhbox{$301$}\\
\hline
 \vcenteredhbox{\includegraphics[scale=\picscalefactor]{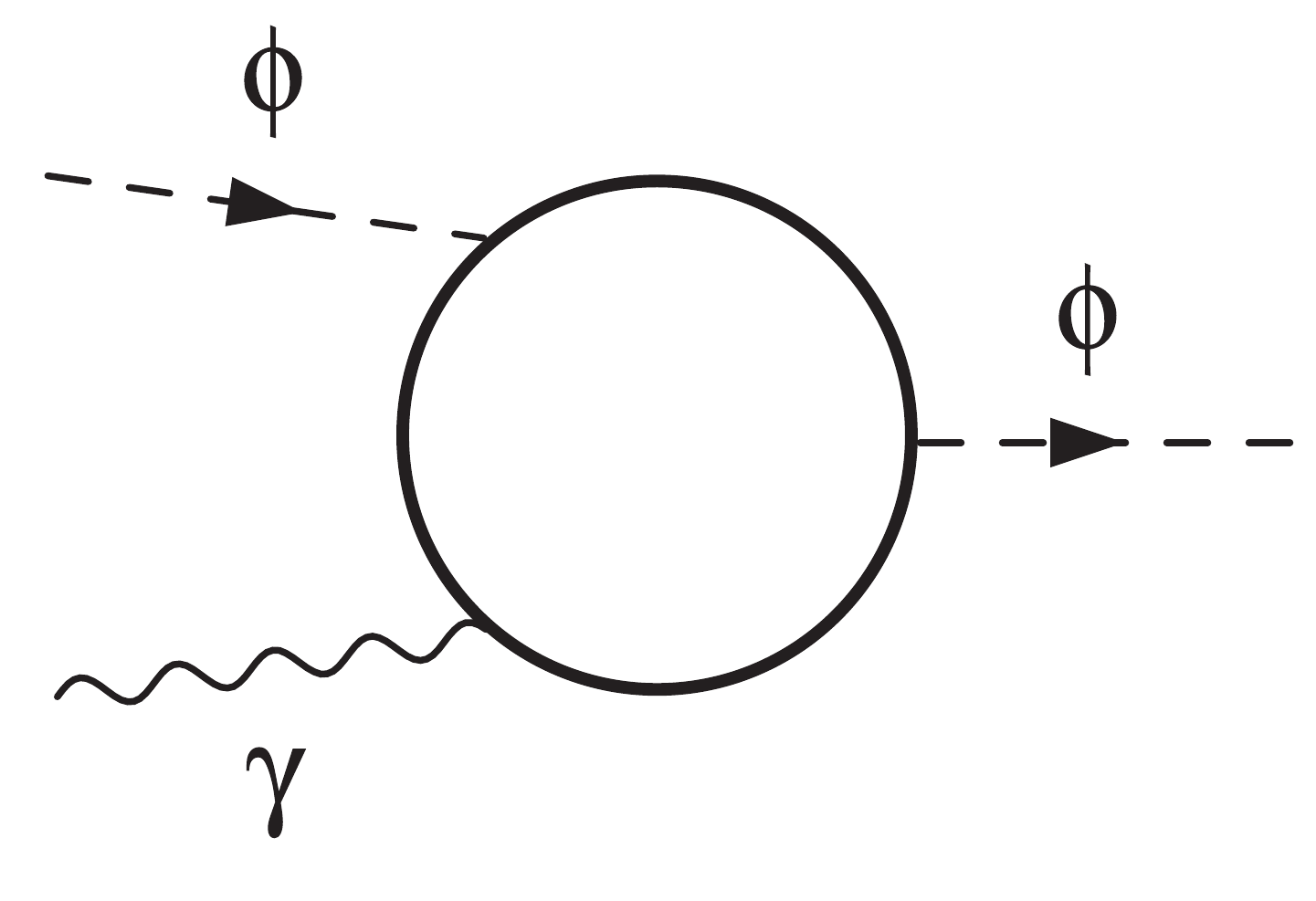}} & \vcenteredhbox{$5$} & \vcenteredhbox{$107$} & \vcenteredhbox{$3084$}\\
\hline
 \vcenteredhbox{\includegraphics[scale=\picscalefactor]{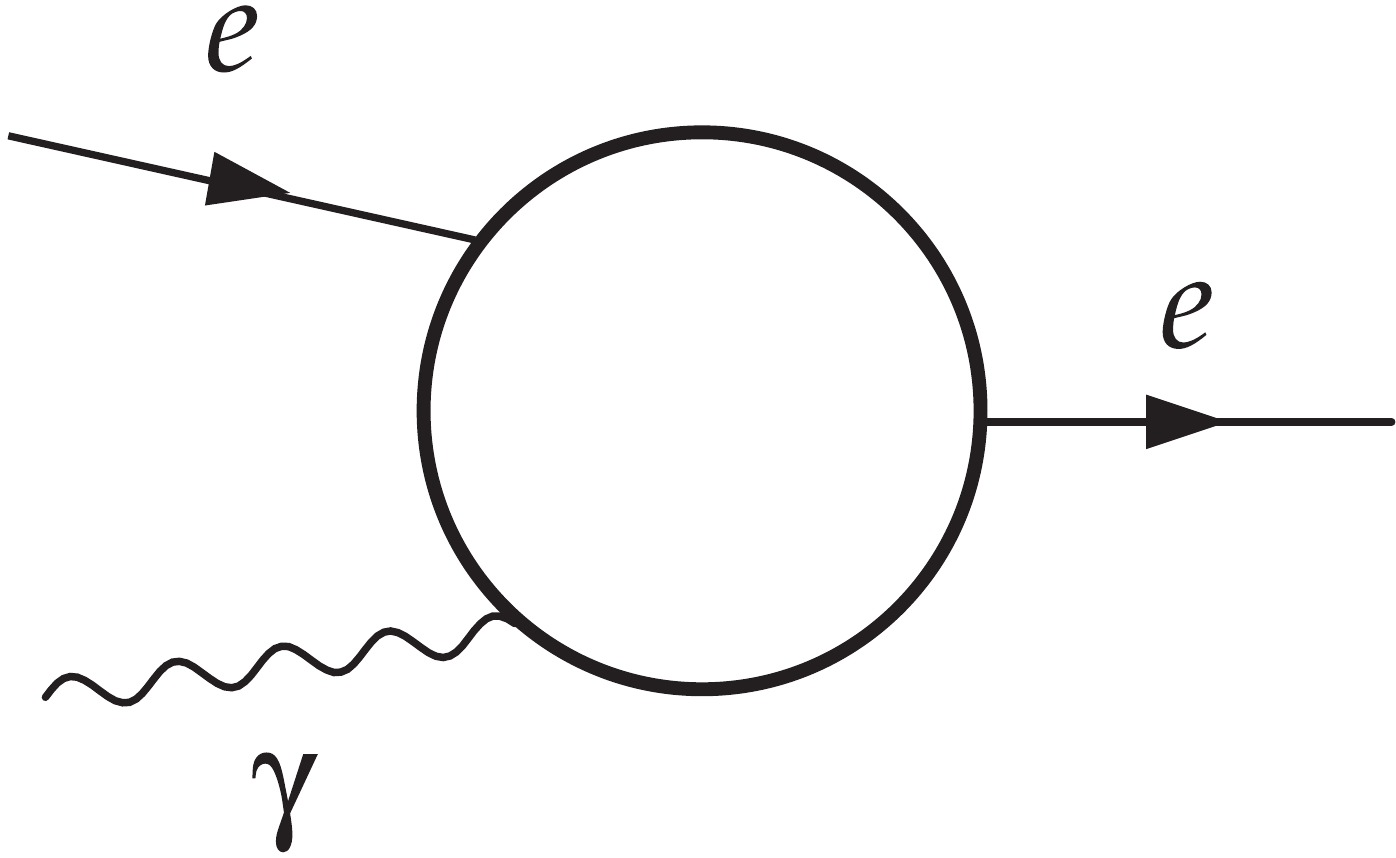}} & \vcenteredhbox{$3$} & \vcenteredhbox{$69$} & \vcenteredhbox{$1996$}\\
\hline
 \vcenteredhbox{\includegraphics[scale=\picscalefactor]{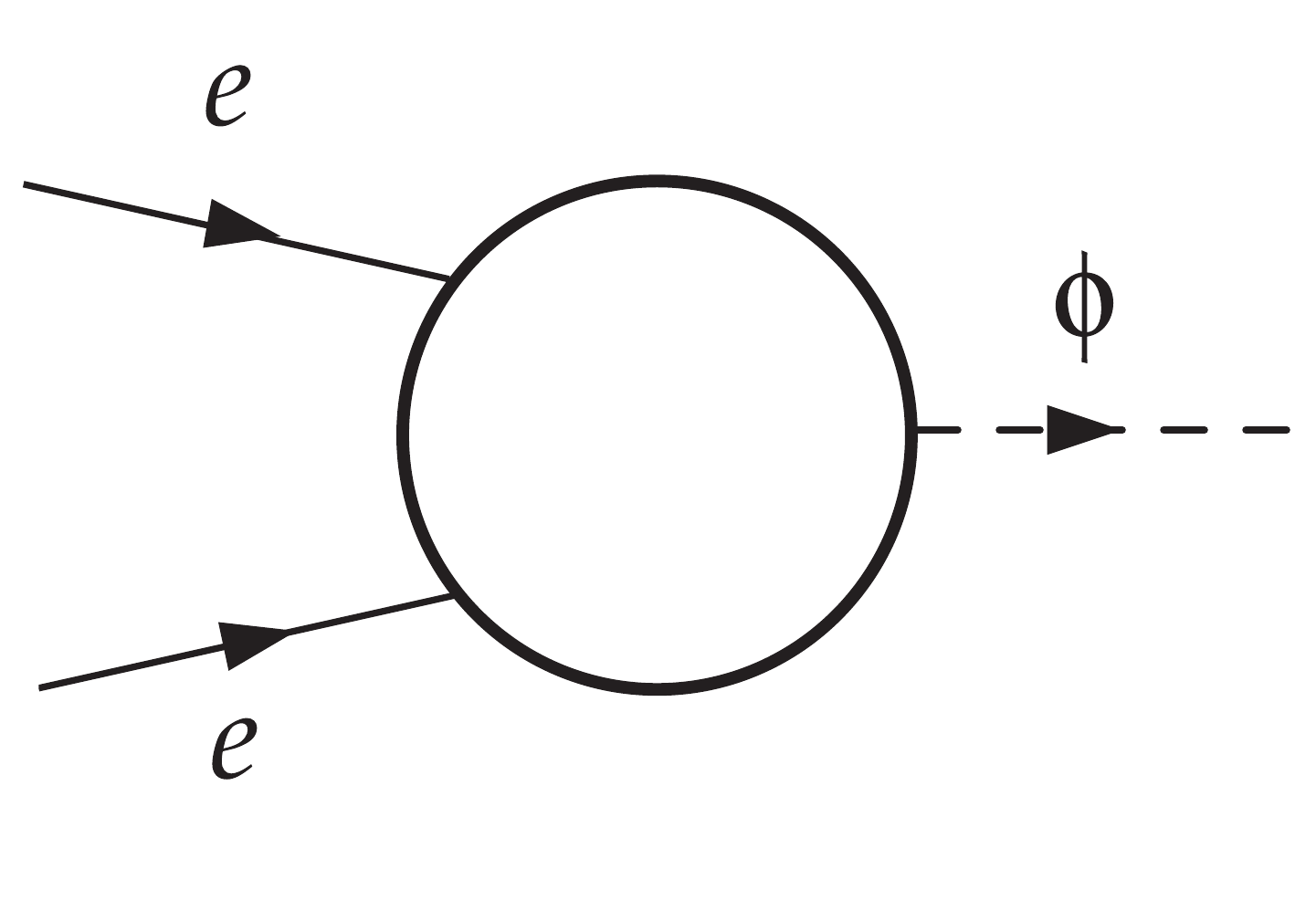}}  & \vcenteredhbox{$3$} & \vcenteredhbox{$64$} & \vcenteredhbox{$1814$}\\
\hline
 \vcenteredhbox{\includegraphics[scale=\picscalefactor]{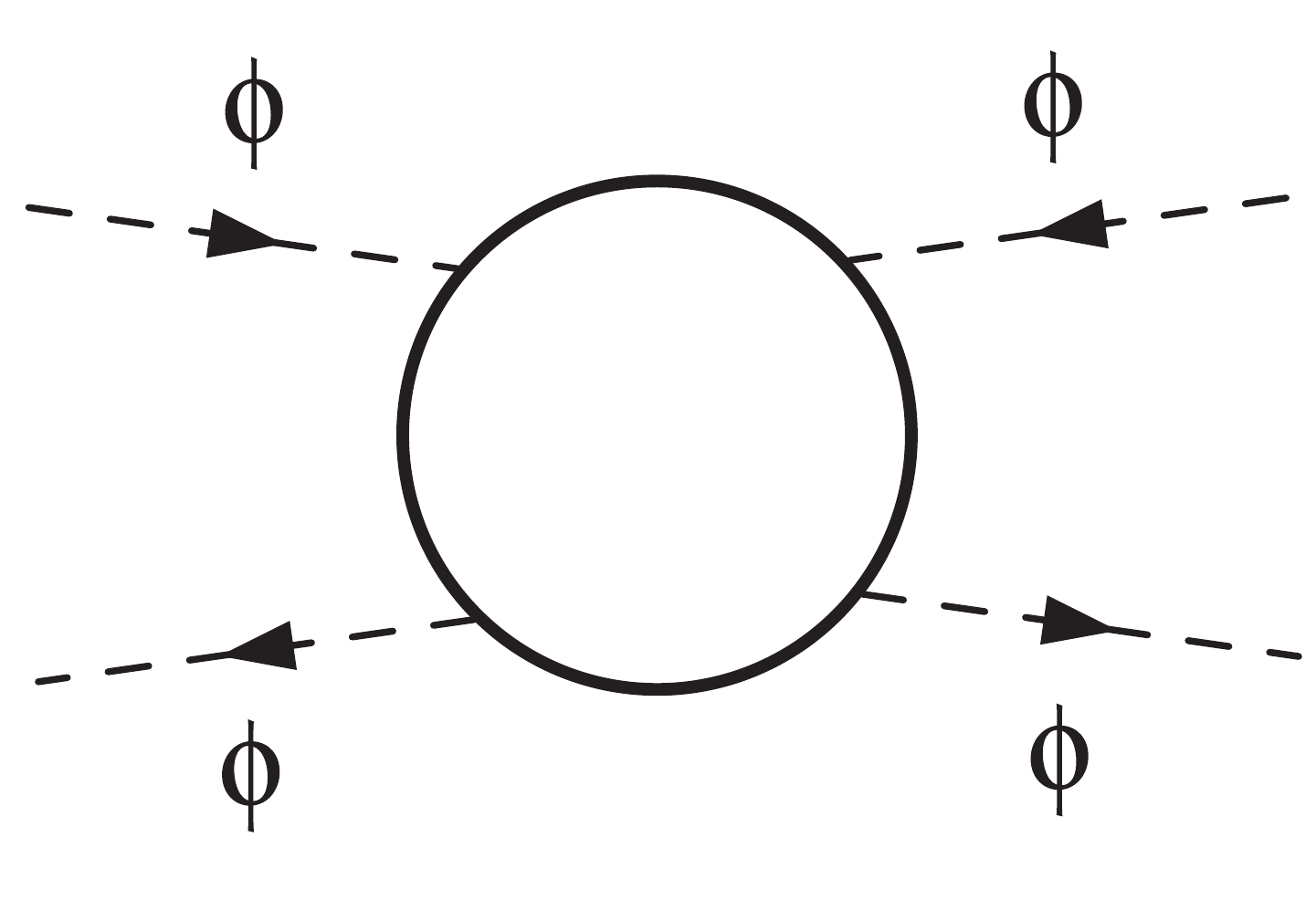}} & \vcenteredhbox{$20$} & \vcenteredhbox{$683$} & \vcenteredhbox{$26961$}\\
\end{tabular}
 \caption{List of all relevant $n$-point functions and the associated number of Feynman diagrams in dependence with the number of loops ($\gamma$: gauge field, $\phi$: Cooper pair field, $e$: Dirac fermion).\label{TAB:DiagramListing}}
\end{table}

In this Appendix we provide some technical details concerning the calculation of the renormalization constants of the theory (\ref{LGwithA}) at three-loop order. In Table~\ref{TAB:DiagramListing} we show the various $n$-point functions which need to be evaluated in order to obtain the required renormalization constants.
Further we list the number of Feynman diagrams in dependence with the number of loops.
From the table it becomes clear that starting at two-loop level the large number of Feynman diagrams cannot be calculated by hand, even if one is only interested in the UV divergent parts of the diagrams.
That is why we rely on a full automated calculation.

In order to generate the complete set of Feynman diagrams we use the program \textsc{QGRAF}~\cite{Nogueira:1991ex}.
The output of \textsc{QGRAF} is further processed by the programs \textsc{q2e} and \textsc{exp}~\cite{Harlander:1997zb,Seidensticker:1999bb} which map all diagrams onto massive tadpole integrals expanding naively in (if required) small external momenta.
Using projectors to obtain the coefficients of the relevant Lorentz structures we are left with the calculation of traces over Dirac matrices and fully contracted Lorentz tensors.
This task is performed by \textsc{FORM}~\cite{Vermaseren:2000nd,Kuipers:2012rf}.
In a last step the reduction to master integrals is performed via integration-by-parts relations by the \textsc{FORM} package \textsc{MATAD}~\cite{Steinhauser:2000ry}.
The latter can deal with one-scale massive tadpole integrals up to and including three loops.

In order to prevent dimensional regularization (DREG) from generating infrared poles in $\epsilon$, we use a common regulator mass $m$ in all propagator denominators.
Because the UV structure of the theory does not depend on this mass and we are allowed to choose the external momenta in a convenient way, 
we can extract the proper UV pole structure in $\epsilon$ just by calculating single-scale massive tadpole integrals after expanding in at most one external momentum.
At one-loop order this procedure is trivial, because one can just discard all terms artificially depending on $m$ ($m$-terms).
At higher-loop orders one has to impose a proper local subtraction of the artificial $m$-terms, because an $m$-term in a sub-diagram/integral can lead to unwanted $m$-free terms in the full diagram/integral. 
The method of subtracting such terms order-by-order in perturbation theory is known as infrared rearrangement;
it was suggested in Ref.~\cite{Misiak:1994zw} and further developed in Ref.~\cite{Chetyrkin:1997fm} (for a short introduction see, for example, Ref.~\cite{Zoller:2014xoa}). In order for infrared rearrangement to be successful one has to subtract all sub-divergences of a given diagram.
Besides the mass subtraction terms already mentioned, which only enter through the scalar and photon propagators~\footnote{There are no subtraction terms for the fermion propagator because of discrete symmetries prohibiting the generation of a fermion mass.}, 
one also has to insert the usual counter-terms contained in the renormalization constants of the theory.

The counter-terms include the field (wave function) renormalization counter-terms stemming from propagators and all vertex counter-terms.
The latter are related to coupling renormalization and field renormalization constants.
In practice one determines the field renormalization constants from the divergent structure of the corresponding two-point function
and vertex renormalization constants from the divergent structure of the corresponding vertex functions. 
This is done iteratively starting at one-loop order.
Here the divergent pieces of the $n$-point function define the corresponding one-loop renormalization constant.
At loop order $L$, one has to insert the expressions already obtained for all counterterms at lower loop orders $0<L_{C}<L$ in all contributing lower loop order ($0<L_{C_d}<L$) counter-term diagrams (with $L=L_{C}+L_{C_d}$)
and add their contribution to the genuine $L$-loop amplitude in order to obtain the remaining $L$-loop divergence which then defines the corresponding $L$-loop counter-term.
The large number of diagrams again requires that the generation of all possible counter-term insertion diagrams is performed automatically.
From a technical point of view this can be easily dealt with by introducing additional vertex labels and the insertion of an explicit counter- and subtraction-term expansion for every propagator.

From the perspective of the automated software packages mentioned previously, one necessary modification arising from the theory (\ref{LGwithA}) is the ability to deal with indefinite fermion flow directions. The Feynman rules for the theory contain two vertices allowing the annihilation of two identical Dirac fermions into a bosonic Cooper pair or the creation of two identical Dirac fermions as decay of the Cooper pair.
This leads to Feynman diagrams where the fermion flow arrows in a single fermion chain can point in opposite directions and thus do not form a unique fermion flow direction; in this case we speak of a fermion flow conflict. However, in all diagrams (including projectors) there is always an even number of fermion flow conflicts. As a result, one can anticommute the $i\sigma_2$ coupling (rewritten as the two-dimensional Levi-Civita or $\varepsilon$-tensor) appearing in the mentioned vertices through the chain to another  $i\sigma_2$ coupling, where they annihilate. This forms chains with a unique, well-defined fermion flow not involving any transposed Dirac structures.
In our \textsc{FORM} implementation we use a naively anticommuting $\gamma_5$ in order to emulate this $SU(2)$-algebra-specific behavior of the two-dimensional $\varepsilon$-tensor with generic Dirac matrices.
By naively anticommuting, we mean here that all $\gamma_5$ matrices are anticommuted through all other Dirac matrices to one side of the fermion trace in order to give unity when we have two of them next to each other.

However, before one can deal with any expression within \textsc{FORM} one needs to generate the Feynman diagrams including the identical fermion vertices.
Although \textsc{QGRAF} provides the correct absolute value for the symmetry factor of all diagrams using identical fermions at one vertex, it does not allow in our setup for an unambiguous definition of the sign of the Wick contraction,
stemming from the anticommuting property of the fermionic field operators.
This is so because the sign stemming from the Wick contraction is correlated with the order of the indices used in the two-dimensional $\varepsilon^{\alpha \beta}$,
but since \textsc{QGRAF} in our setup only works with generic expressions for the vertex involving only the identical names of the connected fields (and thus without explicit dependence on $\alpha$ and $\beta$) we run into a sign ambiguity when working with identical fermions.

To circumvent this problem we first run \textsc{QGRAF} in the identical-particle setup.
Then we simply choose a direction in all fermion chains to be the proper fermion flow direction or Denner current~\cite{Denner:1992me} and replace all fermions creating a propagator with opposite directions by the transposed fermion.
The latter are now effectively a distinct type of fermion, and their propagator direction is opposite to that of normal fermions.
Since we have now two distinguishable fermions at every vertex, we can recalculate the sign of the diagram prefactor stemming from the Wick contraction and thus fix it unambiguously.

Due to the large number of diagrams we need to perform the modification steps above in a fully automated way.
For the reordering of the fermion flow we use the \textsc{PERL} script \textsc{majoranas.pl} written by R. Harlander in order to resolve the indefinite fermion flow direction problem appearing in supersymmetric theories with Majorana fermions~\cite{Harlander:2009mn}.
The recalculation of the signs of the diagram prefactors in the distinguishable fermion setup is done by an in-house \textsc{Mathematica} program.

In order to test our setup including the renormalization routines we reproduced the QED beta function up to three loops.
We explicitly checked our QED result against the Abelian limit of the QCD result presented in the four-loop calculation of Ref.~\cite{vanRitbergen:1997va} (see also Ref.~\cite{Chetyrkin:2004mf,Czakon:2004bu,Zoller:2015tha}).
The three/four-loop QED beta functions were presented before in Ref.~\cite{Chetyrkin:1980pr,Vladimirov:1979zm}/\cite{gorishny1991}. For the respective five-loop results, see Ref.~\cite{kataev2012,baikov2012,luthe2016,baikov2016}.

\end{fmffile}

\bibliography{susyNZ}

\end{document}